\documentclass[11pt, letter,fleqn]{article}
\pdfoutput=1

\usepackage{amsmath, amssymb,amsthm,cite}
\usepackage{graphics,color}
\usepackage{subcaption}
\usepackage{cprotect}
\usepackage{epstopdf}
\usepackage {graphicx}
\usepackage{verbatim}
\usepackage{listings}
\usepackage{tikz}
\usepackage[section]{placeins}
\usepackage{multirow}
\usepackage{longtable}	
\usepackage{pdflscape}	
\usepackage{enumitem}  
\usepackage{IEEEtrantools}
\usetikzlibrary{positioning,calc,cd}

\title{Locally and Globally Optimal Configurations of $N$ Particles on the Sphere with Applications in the Narrow Escape and Narrow Capture Problems }

\author{  Wesley J. M. Ridgway\footnotemark[1], ~~ Alexei
  F. Cheviakov \footnotemark[2]\\ {\small
    \emph{Department of Mathematics and Statistics, University of
      Saskatchewan, Saskatoon, S7N 5E6 Canada}}
}

 {\theoremstyle{definition}

{\theoremstyle{definition} }
{\theoremstyle{definition} }

\newcommand{\x}{x}

\def\max{\mathop{\hbox{\rm max}}}

\def\beq{\begin{equation}}
\def\eeq{\end{equation}}
\def\barr{\begin{array}{ll}}
\def\earr{\end{array}}

\begin{document}

\footnotetext[1]{Electronic mail: wjr704@mail.usask.ca}

\footnotetext[2]{Corresponding author. Electronic mail: cheviakov@math.usask.ca
  }

\maketitle

\begin{abstract}

Determination of \emph{optimal} arrangements of $N$ particles on a sphere is a well-known problem in physics. A famous example of such is the Thomson problem of finding equilibrium configurations of electrical charges on a sphere.  More recently however, similar problems involving other potentials and non-spherical domains have arisen in biophysical systems. Many optimal configurations have previously been computed, especially for the Thomson problem, however few results exist for potentials that correspond to more applied problems. Here we numerically compute optimal configurations corresponding to the \emph{narrow escape} and \emph{narrow capture} problems in biophysics.  

We provide comprehensive tables of global energy minima for $N\leq120$ and local energy minima for $N\leq65$, and we exclude all saddle points.  Local minima up to $N=120$ are available online.  

\end{abstract}

\section{Introduction}
\label{sec:intro}

The problem of optimally distributing $N$ points on the boundary of a bounded domain, $\mathcal{D} \subseteq \mathbb{R}^n$ $(n\geq2)$, has many applications in physical systems (see e.g. Ref. \cite{bowick2009two} and numerous references therein).  Here, \emph{optimal} refers to an arrangement of points, particles, etc. on the boundary, $\partial\mathcal{D}$, such that the configuration of particles $\left\{\mathbf{x}_i\right\}_{i=1}^N$ minimizes the pairwise `potential energy'
\begin{equation}
	\mathcal{H}(\mathbf{x}_1, ..., \mathbf{x}_N)=\sum_{i<j}^N h(|\mathbf{x}_i-\mathbf{x}_j|)
	\label{eqn:general_potential}
\end{equation}
with the constraint $\mathbf{x}_i\in\partial\mathcal{D}$ where $h$ is the interaction energy between two particles.  Such a configuration is called an \emph{optimal configuration}. Note that optimal configurations include local minima of \eqref{eqn:general_potential} as well as the global minimum.  While a large number of algorithms have been used to seek global extrema of general or specific classes of functions \cite{pinter2013global, horst2013handbook, davis1991handbook, ingber1992genetic}, very few algorithms exist for the study of local minima \cite{mehta2016kinetic}.  A large part of this paper is devoted to systematic computations of \emph{local} minima which presents a difficult computational problem due to the number of local minima. In fact, it is believed that the number of local minima increases exponentially \cite{mehta2016kinetic}.

One example of a problem of practicle interest that yields the above-described optimization problem is the \emph{narrow escape problem} in which a Brownian particle diffuses in a bounded domain $\Omega \subset \mathbb{R}^n$ $(n\geq 3)$.  The boundary $\partial\Omega=\partial\Omega_a\cup\partial\Omega_r$ consists of reflecting and absorbing regions, denoted $\partial\Omega_r$ and $\partial\Omega_a$ respectively, where the absorbing regions are small windows or traps with measure $|\partial\Omega_a|=\mathcal{O}(\epsilon^{n-1})$, $0<\epsilon\ll1$.  The narrow escape problem consists of finding the mean first passage time (MFPT), defined as the expectation value of the time required for the Brownian particle to escape $\Omega$ through $\partial\Omega_a$ in the limit where $\epsilon$ is asymptotically small.  The narrow escape problem is a singular perturbation problem since the MFPT diverges in the limit as $\epsilon\rightarrow0$ \cite{pillay2010asymptotic, cheviakov2010asymptotic}.  

\begin{figure}
	\centering
	\includegraphics[width=8cm, height=7cm]{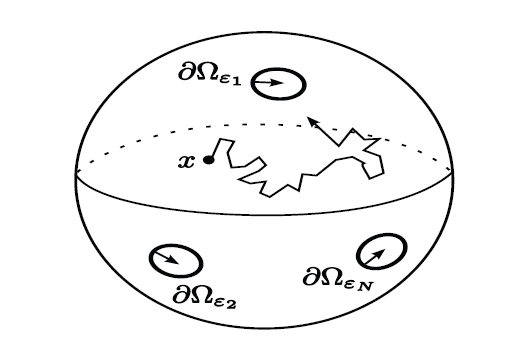}
	\caption{Illustration of Brownian motion in a sphere with absorbing patches on an otherwise reflecting boundary corresponding to the BVP \eqref{eqn:NE_bvp}.  }
	\label{fig:narrow_escape_sphere}
\end{figure}

When $n=3$ and the absorbing boundary consists of $N$ disjoint circular caps of a common (dimensionless) radius $\epsilon$, $\partial\Omega_{\epsilon_i}$, $i=1,...,N$, the MFPT, $v(x)$, satisfies the boundary value problem
\begin{equation}
	\begin{cases}
	\Delta v = -\frac{1}{D}, \qquad x\in \Omega, \\
	v=0, \quad x\in\partial\Omega_a=\bigcup\limits_{i=1}^N\partial\Omega_{\epsilon_i}, \\
	\partial_n v = 0, \quad x\in\partial\Omega_r,
	\end{cases}
	\label{eqn:NE_bvp}
\end{equation}
where $D$ is the diffusivity of the Brownian motion.  The situation is depicted in Figure \ref{fig:narrow_escape_sphere}. We remark that in 2D a heterogeneous Dirichlet-Neumann problem is also known as the \emph{Keldysh-Sedov problem}, originally considered for the Laplace equation (see e.g. Refs \cite{keldysh1937effective, lavrentiev1987methods}). 

The average MFPT for a uniform distribution of starting locations for the Brownian particle is computed from 
\begin{equation}
	\bar{v}= \frac{1}{|\Omega|}\int_\Omega v(x) dx,
	\label{eqn:avg_mfpt}
\end{equation}
where $|\Omega|$ is the volume of $\Omega$.

The narrow escape problem has application in biophysics where proteins, ions, etc. (modelled as Brownian particles) diffuse in a confining domain.  Specific examples include virus transport inside cell nuclei, chemical reactions in microdomains, and the motion of calcium ions in dendritic spines (see Refs. \cite{schuss2007narrow, schuss2015brownian} therein).  As there are usually many Brownian particles, the average MFPT, \eqref{eqn:avg_mfpt}, acts as a timescale for these particles to exit the domain to accomplish some biological function. A natural problem arising out of this model is that of computing arrangements of traps that minimize the average MFPT.  This constitutes a constrained optimization problem in which the objective function is of the form \eqref{eqn:general_potential} and will be discussed in greater detail in the next section.  

\begin{figure}
\begin{tikzpicture}[>=stealth]
	\draw [line width=0.25mm](0,0) .. controls (-1,3) and (3,3) .. (5,3);
	\draw [line width=0.25mm](5,3) .. controls (6,3) and (7,3) .. (7,1);
	\draw [line width=0.25mm](7,1) .. controls (7,0) and (6,-1) ..  (5,-1.5);
	\draw [line width=0.25mm](5,-1.5) .. controls (4,-2) and (1,-3) .. (0,0);
	\node at (4,2){$\Omega$};

	\draw (5,0) circle (0.5);
	\draw[dashed] (4.5,0) to [bend left](5.5,0);
	\draw (4.5,0) to [bend right](5.5,0);
	\draw (5,0) -- (4.25,0.75) node[shift={(-0.15,0.15)}]{$\mathbf{x}_0$};
	\draw (5,-0.25) -- (4.5,-0.75) node[shift={(-0.15,-0.15)}]{$\Omega_\epsilon$};
	\draw[fill] (5,0) circle (0.05);

	\draw[dashed] (5,0.5) -- (10.3,2.4);
	\draw[dashed] (5,-0.5) -- (10.3,-2.4);

	\draw [line width=0.25mm](11,0) circle (2.5);
	\draw [line width=0.25mm](8.5,0) to[bend right](13.5,0);
	\draw[line width=0.25mm, dashed] (8.5,0) to[bend left](13.5,0);

	\draw[line width=0.5mm](11, 2) ellipse (0.4 and 0.3);
	\draw[line width=0.5mm, rotate around={-30:(12.5,-1)}](12.5,-1.25) ellipse(0.3 and 0.45);
	\draw[line width=0.5mm, dashed, rotate around={30:(9.5,-1)}](9.5,-1) ellipse(0.3 and 0.45);

	\node at (14,2){$\partial\Omega_{\epsilon_a}$};
	\draw[->] (13.5,2) -- (11.5,2);
	\draw[->] (13.5,2) -- (12.5,-0.7);
	\draw[->] (13.5,2) -- (9.9,-1);

	\draw[ line width=0.25mm] (1,2)
	\foreach \x in {1,...,121}
	{ -- ++(rand*0.3,rand*0.3)
	};
	\draw[line width=0.25mm, ->] (2.62,-0.28)--(3,0);
	\draw[fill] (1,2) circle (0.05) node[above]{$\mathbf{x}$};
\end{tikzpicture}
	\caption{Illustration of the narrow capture problem consisting of a single target sphere with absorbing patches inside a general domain.}
	\label{fig:narrow_capture_sphere}
\end{figure}
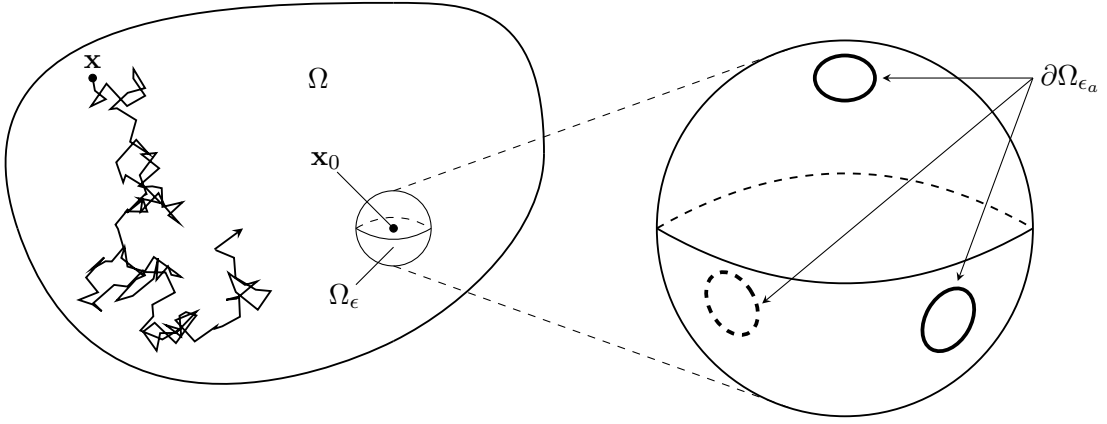

Another specific example is the \emph{narrow capture problem}, similar to the narrow escape problem, which also has applications in biophysics (see Refs. \cite{lindsay2015narrow, lindsay2017first} and references therein).  Here, there are $M$ disjoint interior absorbing traps, $\Omega_\epsilon \subset \Omega$, each with a size parameter $\epsilon \ll 1$, and the boundary of $\Omega$ is entirely reflecting \cite{cheviakov2011optimizing}.  The case where $\Omega \subset \mathbb{R}^3$ contains a single spherical trap ($M=1$) of radius $\epsilon$ centered at $\mathbf{x}_0\in\Omega$ was studied in Ref. \cite{lindsay2017first}.  The boundary of the target sphere is reflecting except for $N$ disjoint circular absorbing nanotraps of a common radius, $\sigma$. That is, $\partial\Omega_\epsilon=\partial\Omega_{\epsilon_r}\cup\partial\Omega_{\epsilon_a}$ where $\partial\Omega_{\epsilon_a}$ and $\partial\Omega_{\epsilon_r}$ are the absorbing and reflecting regions respectively.  The situation is depicted in Figure \ref{fig:narrow_capture_sphere}. The MFPT $v(x)$ for a Brownian particle starting at $x\in\Omega \backslash\Omega_\epsilon$, satisfies the boundary-value problem
\begin{equation}
	\begin{cases}
	\Delta v = -\dfrac{1}{D}, \qquad x\in\Omega\backslash\Omega_\epsilon, \\
	v = 0, \qquad x \in \partial\Omega_{\epsilon_a}, \\
	\partial_n v = 0, \qquad x\in \partial\Omega \cup \partial\Omega_{\epsilon_r}.
	\end{cases}
\label{eqn:NC_bvp}
\end{equation}
The average MFPT is computed using \eqref{eqn:avg_mfpt} but replacing $\Omega$ with $\Omega\backslash\Omega_\epsilon$.  

\pgfmathsetseed{2}

The narrow capture problem models some biophysical processes.  For example, a cell transports proteins (again modelled by Brownian particles) between the cytoplasm and the cell nucleus which is roughly spherical in shape.  Proteins are allowed to pass through the otherwise impermeable nuclear membrane via nuclear pore complexes (NPCs) distributed over the surface.  There are near 2000 of these pores covering about 2\% of the nuclear surface, each has a radius of about 25nm \cite{lindsay2017first}.  The average MFPT again provides a timescale for the large number of particles to escape the domain to accomplish a biological function.   

Another practical problem is modelling the diffusion of nanoparticles in so-called inverse opals.  Inverse opals are porous materials consisting of connected spherical cavities.  These materials have several applications in engineering and are interesting from a material science perspective (see e.g. Refs. \cite{raccis2011confined, skaug2015tracking} and references therein).  Although diffusion in these structures has been studied experimentally and numerically (see e.g. Ref. \cite{cherdhirankorn2010tracer}), development of a general model of the diffusion process that takes into account the geometric structure is an open problem.  The MFPT for a spherical cavity with pores may provide a measure of the average `dwell' time of a particle in each cavity.  One would expect optimal configurations to maximize the diffusion which is desirable when designing materials for industrial processes.  Thus the narrow escape problem and corresponding optimal configurations are relevant in the modelling of this problem.

Optimal configurations on the sphere in $\mathbb{R}^3$ for the general pairwise potential \eqref{eqn:general_potential} where $h$ in \eqref{eqn:general_potential} is a continuous and monotonically decreasing function have previously been studied (e.g. Refs. \cite{cohn2007universally, ballinger2009experimental}).  It is well known that the optimal configurations depend on the potential for all $N$ with a few exceptions ($N=2,3,4,6,12$).  The global mininima for these $N$ are optimal for every such $h$ and these configurations are termed \emph{universally optimal}.  The universal optima in $\mathbb{R}^3$ consist of antipodal points ($N=2$), an inscribed equilateral triangle ($N=3$), a tetrahedron ($N=4$), an octahedron ($N=6$), and an icosahedron ($N=12$).  This is a complete list for $\mathbb{R}^3$ \cite{ballinger2009experimental}.

It is also well known that it is geometrically impossible, in general, to evenly distribute points on a sphere.  If $c_i$ is the coordination number (number of nearest neighbours) of particle $i$, then the Euler's formula for the sphere yields (e.g. \cite{bowick2009two})
\begin{equation}
	\sum_{i=1}^N(6-c_i) = 12,
	\label{eqn:sphere_defect}
\end{equation}
which shows that the total defect structure must always be 12. For example, a tessellation of the sphere with hexagons and pentagons must always contain exactly 12 pentagons.  Our results are consistent with these facts.  In light of equation \eqref{eqn:sphere_defect}, we define a \emph{scar} as a collection of adjacent defects. 

In this paper we compute local and global minima of some selected potentials that have applications in the narrow escape and narrow capture problems up to $N=120$. The rest of the paper is organized as follows: first we summarize the results of Refs. \cite{cheviakov2010asymptotic, lindsay2017first} and present the potentials that we will consider (Section \ref{sec:optimization_problems}). Then in Section \ref{sec:numerical_computation} we describe briefly the optimization algorithm \cite{ridgway2018iterative}.  The remaining sections present the results with comprehensive tables of local and global minima as well as interpretation and comparison with previous work \cite{cheviakov2013narrow}.  

\section{Asymptotic Formulas for the Average MFPT and Related Optimization Problems}
\label{sec:optimization_problems}

The narrow escape problem for the unit sphere with $N$ identical `circular' traps of radius $\epsilon << 1$ was studied in  \cite{cheviakov2010asymptotic}.  The MFPT was computed using the method of matched asymptotic expansions using information about the surface Neumann Green's function for the sphere.  It was shown that the average MFPT has the following three-term expansion:

\begin{equation}
	\bar{v} = \frac{|\Omega|}{4\epsilon D N}\left[ 1 + \frac{\epsilon}{\pi}\log\left(\frac{2}{\epsilon}\right)  + \frac{\epsilon}{\pi}\left( -\frac{9N}{5} + 2(N-2)\log 2 + \frac{3}{2} + \frac{4}{N}\mathcal{H}_\text{NE}(x_1, ..., x_N)\right) + \mathcal{O}(\epsilon^2 \log\epsilon) \right]
\end{equation}
where $D$ is the diffusivity of the Brownian motion, $|\Omega|$ is the volume of the unit sphere, and the function $\mathcal{H}_\text{NE}(x_1,...,x_N)$ is a discrete pairwise energy given by
\begin{equation}
	\mathcal{H}_\text{NE} = \sum_{i<j}^N\left( |\mathbf{x}_i-\mathbf{x}_j|^{-1} - \frac{1}{2}\log|\mathbf{x}_i-\mathbf{x}_j|    -  \frac{1}{2}\log(2+ |\mathbf{x}_i - \mathbf{x}_j|)\right).
	\label{eqn:ne_energy}
\end{equation}
The first two terms are the usual Coulombic and logarithmic potentials respectively
\begin{equation}
	\mathcal{H}_\text{C}	= \sum_{i<j}^N|\mathbf{x}_i-\mathbf{x}_j|^{-1},
	\label{eqn:coul_energy}
\end{equation}
\begin{equation}
	\mathcal{H}_\text{L}	=-\sum_{i<j}^N\log|\mathbf{x}_i-\mathbf{x}_j|. 
	\label{eqn:log_energy}
\end{equation}
Thus the configuration of traps that minimizes the MFPT is obtained by minimizing the discrete pairwise energy \eqref{eqn:ne_energy}.

The narrow capture problem for $M$ interior targets was analyzed in Ref. \cite{cheviakov2011optimizing} using the method of matched asymptotic expansions in order to arrive at a three-term expansion in $\epsilon$ for the MFPT.  For a single target ($M=1$) the result for the average MFPT reduces to 
\begin{equation}
	\bar{v}=\frac{|\Omega|}{4\pi C D\epsilon}\left[ 1 +4\pi C R(\mathbf{x}_0) + \mathcal{O}(\epsilon^2)\right]
\end{equation}
where $R(\mathbf{x}_0)$ is the regular part of the Neumann Green's function for $\Omega$.  The quantity $C$, called the capacitance of the target sphere, is defined in terms of an electrostatic potential problem.  When the target sphere consists of a reflecting boundary with absorbing patches as in \cite{lindsay2017first}, the capacitance is defined in terms of the boundary-value problem
\begin{equation}
	\begin{cases}
		\Delta \phi = 0, \qquad \mathbf{y}\in \mathbb{R}^3\backslash \mathcal{B}, \\
		\phi=0, \qquad \mathbf{y}\in \Gamma_a, \\
		\partial_n \phi=0, \qquad \mathbf{y}\in\Gamma_r, \\
		\displaystyle\lim\limits_{R\rightarrow\infty}\int\limits_{\partial\mathcal{B}_R}\partial_n\phi dS= -4\pi,
	\end{cases}	
\end{equation}
where $\mathcal{B}$ is the magnified target sphere centered at the origin with unit radius and $\mathcal{B}_R$ is a sphere of radius $R$ centered at the origin.  The reflecting and absorbing areas on $\mathcal{B}$ are denoted by $\Gamma_r$ and $\Gamma_a$ respectively.  The far-field behaviour of the solution defines $C$ by
\begin{equation}
	\phi \sim \frac{1}{|\mathbf{y}|}-\frac{1}{C}\left[ 1-\frac{\mathbf{p}\cdot\mathbf{y}}{|\mathbf{y}|^3} \right] +... \qquad\text{as } |\mathbf{y}|\rightarrow\infty
\end{equation}
where $\mathbf{p}$ is the dipole moment corresponding to the magnified target sphere (c.f. Ref. \cite{cheviakov2011optimizing}).  In Ref. \cite{lindsay2017first} it was shown that the capacitance for the target sphere is given by 
\begin{equation}
	\frac{1}{C}=\frac{\pi}{N\sigma}\left[ 1+\frac{\sigma}{\pi}\left( \log\left( 2e^{-3/2\sigma}\right)+\frac{4}{N}\mathcal{H}_\text{NC} \right) +\mathcal{O}(\sigma^2\log\left(\frac{\sigma}{2}\right)\right].
\end{equation}
As is the case with the MFPT for the narrow escape problem, $\mathcal{H}_\text{NC}$ is a discrete energy-like function defined by
\begin{equation}
	\mathcal{H}_\text{NC}= \mathcal{H}_\text{C} - \frac{1}{2}\mathcal{H}_\text{L} - \sum_{i<j}^N \frac{1}{2}\log(2+ |\mathbf{x}_i - \mathbf{x}_j|).
	\label{eqn:nc_energy}
\end{equation}
Minimizing $\mathcal{H}_\text{NC}$ minimizes the average MFPT for the narrow capture problem.  Equations \eqref{eqn:nc_energy} and \eqref{eqn:ne_energy} differ only in the sign of $\mathcal{H}_\text{L}$.

The main focus of this paper is to systematically compute optimal configurations of particles on the sphere that minimize \eqref{eqn:ne_energy} and \eqref{eqn:nc_energy} when $N\leq 120$.  We will refer to these potentials as the narrow escape (NE) and narrow capture (NC) potentials respectively.  Optimal configurations for the narrow escape potential have previously been computed, but only the global minima \cite{cheviakov2010asymptotic, cheviakov2013narrow}.  In \cite{cheviakov2012mathematical} optimal configurations were computed for the narrow escape potential with differently 'charged' particles.  Here we attempt to compute all local minima for $N$ identically charged particles in addition to the global minimum.  The optimal configurations for the narrow capture potential have not previously been studied.  

Optimal configurations for more common potentials, such as the Coulombic \eqref{eqn:coul_energy} and logarithmic \eqref{eqn:log_energy} potentials, have been studied more extensively \cite{ridgway2018iterative, erber1991equilibrium, erber1995comment, erber1996complex, bergersen1994equilibrium}.  These potentials also have several physical applications and are used in benchmarking optimization software.  For example, the determination of optimal configurations on the sphere for the Coulomb potential \eqref{eqn:coul_energy} constitutes the classic and well-known Thompson problem.  The modelling of multi-electron bubbles in liquid helium is a modern example of the Thompson problem \cite{guo2009theory}.  The logarithmic potential has applications in the modelling of vortex defects in superconductors \cite{bowick2009two}.

\begin{figure}[htb!]
	\centering
	\includegraphics[width=0.7\textwidth]{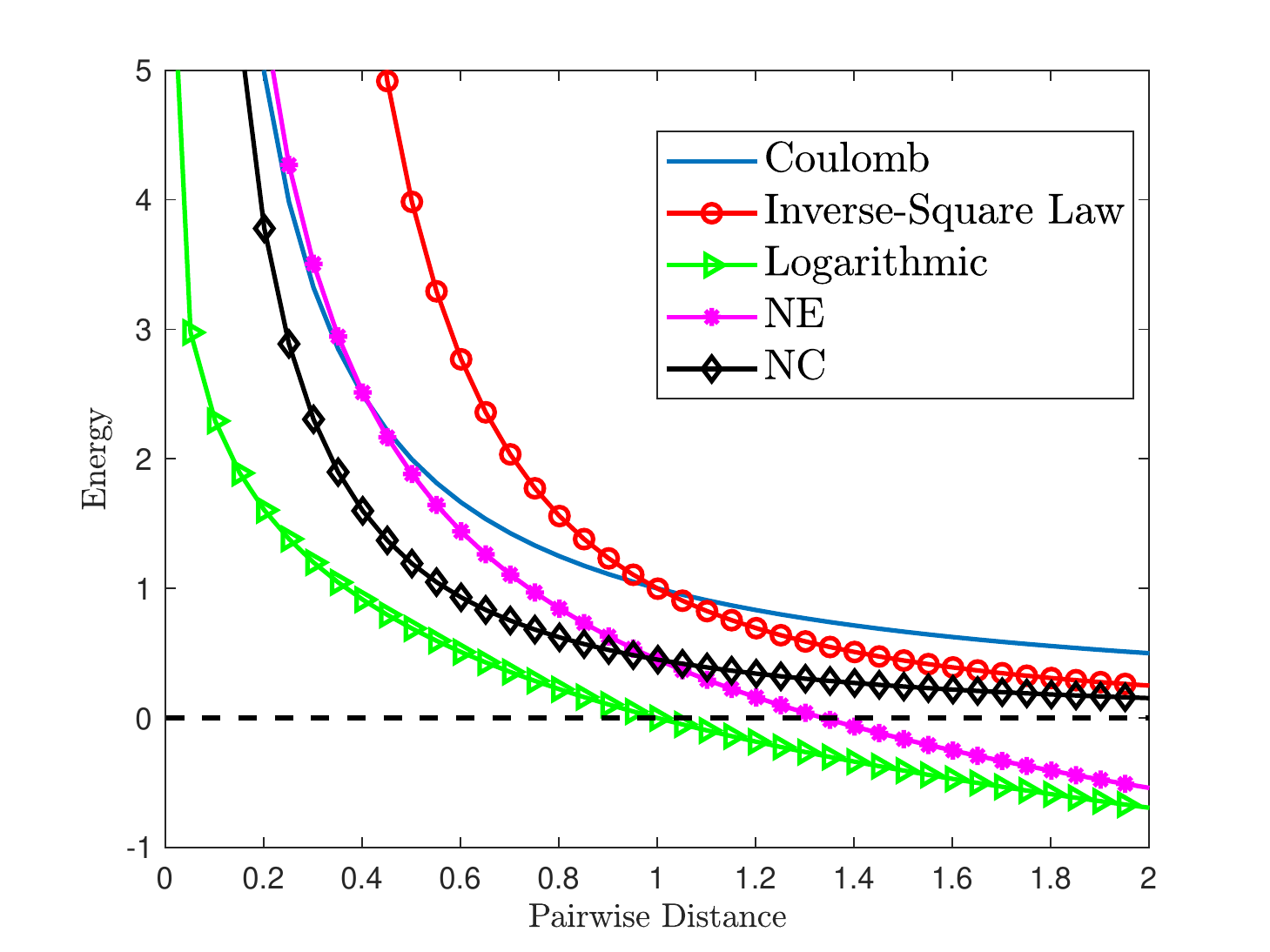}
	\caption{Pairwise energies for each potential \eqref{eqn:ne_energy} - \eqref{eqn:log_energy}, \eqref{eqn:nc_energy}, and \eqref{eqn:inverse_power_energy}. Over the unit sphere, the pairwise distance is at most 2. (Color online).}
	\label{fig:potentials}
\end{figure}

Optimal configurations for inverse power law potentials relate closely to packing problems. Finding the most efficient packing of spherical caps on the surface of a sphere constitutes the best-packing problem for the sphere, also known as the Tammes problem.  The solutions to this problem are given by optimal configurations of a short range power law
\begin{equation}
	\mathcal{H}_m=\sum_{i<j}^N\frac{1}{|\mathbf{x}_i-\mathbf{x}_j|^m}, \qquad m\rightarrow\infty. 
	\label{eqn:inverse_power_energy}
\end{equation}
The case where $m=2$ was examined numerically in Ref. \cite{ridgway2018iterative}.  Figure \ref{fig:potentials} compares various potential functions.  

\section{Numerical Computation of Optimal Configurations}
\label{sec:numerical_computation}

In this section, we describe briefly the optimization algorithm employed to compute putatively optimal configurations.  Details are given in Ref. \cite{ridgway2018iterative}.  The algorithm consists broadly of three steps.  
\begin{enumerate}
	\item Generation of initial configurations as starting points for optimization
	\item Energy minimization via modified steepest descent
	\item Removal of saddle-points
\end{enumerate}

The algorithm generates $N$-particle starting configurations by computing a triangulation of previously known $(N-1)$-particle optimal configurations.  The $N^{\text{th}}$ particle is inserted at the center of mass of one of the triangles on the convex hull of the triangulation and projected onto the surface of the sphere.  This procedure is performed for each triangle center, thus for each triangle center one obtains a starting configuration.  Due to the possible symmetry of the $(N-1)$-particle configuration, some of the resulting starting configurations may be identical due to rotational and reflection invariance of the energy.  The redundant configurations are identified and excluded by calculating pairwise distances between particles.    

Local optimization is accomplished by a modified steepest descent algorithm.  Define forces, acting on particle $i$ according to 
\begin{equation}
	\mathbf{F}_i=-\nabla_i\mathcal{H}(\mathbf{x}_1,...,\mathbf{x}_N),
	\label{eqn:forces}
\end{equation}
where $\nabla_i$ is the gradient operator with respect to the coordinates of particle $i$.  At each step of the energy minimization, the position of the $i^{\text{th}}$ particle, $\mathbf{x}_i$ is updated according to 
\begin{equation}
	\mathbf{x}_i\rightarrow\frac{\mathbf{x}_i-\gamma\mathbf{F}_i^\tau}{|\mathbf{x}_i-\gamma\mathbf{F}_i^\tau|},
	\label{eqn:descent}
\end{equation}
where $\mathbf{F}_i^\tau$ is the component of $\mathbf{F}_i$ in the tangential direction and $\gamma$ is a constant given by
\begin{equation}
	\gamma=\frac{\beta a_0}{F_{init}^\tau}.
\end{equation}
The largest tangential force in the starting configuration is given by ${F_{init}^\tau}$ and $a_0$ is a characteristic distance related to the number of particles and is approximately half the distance between particles for large $N$.  The parameter $\beta$ is user-specified and here is chosen to be 0.5 based on several trial runs for smaller $N$.  

Occasionally the algorithm will find configurations which are saddle points.  Remarkably, this is a common occurrence for this problem (cf. \cite{erber1991equilibrium, erber1996complex, bergersen1994equilibrium}).  The Hessian matrix in the neighbourhood of these saddle points often has at least one small negative eigenvalue which slows the steepest descent procedure.  In principle this can be solved by taking a larger descent step (choosing a larger $\beta$).  However, the unstable directions are believed to be quite `narrow' such that the forces have only a small component along the unstable directions.  If the optimization algorithm is continued (ignoring the stopping condition) then it eventually finds the local minimum, however doing so is inefficient.  Instead, the putative optimal configurations are tested by computing the eigenvalues of the Hessian matrix and excluding those with negative eigenvalues. Note that the Hessian matrix is semi-definate at every local minima due to the rotational and reflectional invariance.  Thus we fix the position of one particle and the azimuth of a second when computing the Hessian.  

Due to the symmetry of the optimal configurations, many of the generated starting configurations, and putative minima, are identical up to rotation and reflection.  Thus we require a method to distinguish configurations to avoid unnecessary computation and to avoid errors in counting the number of minima.  In principle, one can simply compute the energy of each configuration using any `potential energy' function.  However, for a given $N$ the optimal energies are typically very similar (see figures \ref{fig:spectrum_ne} and \ref{fig:spectrum_nc} in the next two sections).  When $N\leq65$ this doesn't pose a problem (see Ref. \cite{ridgway2018iterative}).  As $N$ increases it becomes more difficult to distinguish configurations using energy. Further, the energy is relatively insensitive to small changes in particle positions around local minima making it difficult to resolve differences between two configurations.

Identifying equivalent configurations is accomplished with the set of $N(N-1)/2$ pairwise distances, which is invariant under rotation and reflection.  Suppose we have $k$ configurations, let $\mathbf{x}_i^{(k)}$ denote the $i^\text{th}$ particle in the $k^\text{th}$ configuration. The algorithm proceeds as follows:
\begin{enumerate}
	\item Compute scalars $d_{ij}^k=|\mathbf{x}_i^{(k)}-\mathbf{x}_j^{(k)}|$, $i<j\leq N$, and let these be the components of $\mathbf{d}^{(k)}$. 
	\item For each $k$, sort $\mathbf{d}^{(k)}$ in ascending order and then normalize such that $\max\limits_{k}\left({d^{(k)}_{ij}}\right)=1$ for a given $i$ and $j$.  Denote the resulting vector by $\tilde{\mathbf{d}}^{(k)}$.
	\item Compute the clustering tolerance
		\begin{equation}
			\delta = \frac{|\emph{tol}|}{ \max\limits_{k}||\tilde{\mathbf{d}}^{(k)}||_{L^2}},
		\end{equation}	
		where $\emph{tol}$ is user specified and was chosen as $10^{-3}$ since this was found to give good results. 
	\item Cluster the $\tilde{\mathbf{d}}^{(k)}$ using Euclidean distance and the tolerance in the previous step.  This was implemented with the Statistics Toolbox in MATLAB.
\end{enumerate}
All configurations within a cluster are identical up to rotation and reflection when an appropriate tolerance is chosen.   

\section{Results for the NE Potential up to $N = 120$ }
\label{sec:results_ne}

Putative globally optimal configurations on the sphere for the NE potential have been computed for $N<65$ by \cite{cheviakov2010asymptotic, cheviakov2013narrow}. In the latter study, globally optimal configurations for a few selected $N$ up to 1004 were found.  Locally optimal configurations have not previously been computed to the authors' knowledge.  The current section is divided into three parts: First we present tables ofthe computed global minima and the corresponding energies.  Then we give results for local minima, including energy spectra, tables up to $N=65$, and scars.  Finally, we compare the globally minimal energies with a previously derived asymptotic scaling law \cite{cheviakov2010asymptotic}.  Data on the local minima for $N> 65$ are available online but are not included here due to the amount of data (see the online description).

\subsection{Global Minima}

Table \ref{table:ne_global} gives the computed globally optimal energies for each $N$ along with the number of computed local minima. Each configuration was verified by computing the eigenvalues of the Hessian matrix.

The computation time for each $N$ increases rapidly due to the increase in local minima.  The majority of the computation time was spent optimizing configurations for $115\leq N \leq 120$ for which the number of starting configurations was between 25000 and 66000.

\begin{center}
\begin{longtable}{|r|r|r|r|r|r|}
\hline
$N$ & NE Energy & \# of Local Minima & $N$ &  NE Energy & \# of Local Minima \\
\hline
4 & -1.6671799 & 1 & 63 & 311.6558511 & 1 \\
\hline
5 & -2.0879876 & 1 & 64 & 324.0896299 & 1 \\
\hline
6 & -2.5810055 & 1 & 65 & 336.7697094 & 1 \\
\hline
7 & -2.7636584 & 1 & 66 & 349.6565931 & 2 \\
\hline
8 & -2.9495765 & 1 & 67 & 362.7514093 & 1 \\
\hline
9 & -2.9764336 & 1 & 68 & 376.2377825 & 3 \\
\hline
10 & -2.8357352 & 1 & 69 & 389.9300126 & 4 \\
\hline
11 & -2.4567341 & 1 & 70 & 403.8308809 & 5 \\
\hline
12 & -2.1612842 & 1 & 71 & 418.0224191 & 1 \\
\hline
13 & -1.3678269 & 1 & 72 & 432.3019807 & 4 \\
\hline
14 & -0.5525928 & 1 & 73 & 447.2100228 & 2 \\
\hline
15 & 0.4774376 & 1 & 74 & 462.2113780 & 9 \\
\hline
16 & 1.6784049 & 2 & 75 & 477.3635907 & 3 \\
\hline
17 & 3.0751594 & 1 & 76 & 492.8366797 & 6 \\
\hline
18 & 4.6651247 & 1 & 77 & 508.4749290 & 4 \\
\hline
19 & 6.5461714 & 1 & 78 & 524.4487720 & 4 \\
\hline
20 & 8.4817896 & 1 & 79 & 540.7371289 & 5 \\
\hline
21 & 10.7013196 & 1 & 80 & 557.2315390 & 7 \\
\hline
22 & 13.1017418 & 2 & 81 & 574.1035388 & 7 \\
\hline
23 & 15.8212821 & 1 & 82 & 591.1522915 & 13 \\
\hline
24 & 18.5819815 & 1 & 83 & 608.4133589 & 16 \\
\hline
25 & 21.7249125 & 1 & 84 & 625.9597981 & 17 \\
\hline
26 & 25.0100312 & 1 & 85 & 643.7723346 & 10 \\
\hline
27 & 28.4296992 & 1 & 86 & 661.8438907 & 22 \\
\hline
28 & 32.1929330 & 1 & 87 & 680.1575495 & 20 \\
\hline
29 & 36.2197826 & 1 & 88 & 698.7043251 & 18 \\
\hline
30 & 40.3544394 & 1 & 89 & 717.5605772 & 15 \\
\hline
31 & 44.7576167 & 1 & 90 & 736.6531884 & 24 \\
\hline
32 & 49.2409494 & 2 & 91 & 756.0230290 & 23 \\
\hline
33 & 54.2959715 & 1 & 92 & 775.6539262 & 24 \\
\hline
34 & 59.3794885 & 1 & 93 & 795.5664664 & 23 \\
\hline
35 & 64.7367107 & 2 & 94 & 815.6923218 & 32 \\
\hline
36 & 70.2760966 & 1 & 95 & 836.1253560 & 22 \\
\hline
37 & 76.0662374 & 2 & 96 & 856.7795176 & 21 \\
\hline
38 & 82.0802998 & 2 & 97 & 877.7410951 & 8 \\
\hline
39 & 88.3295602 & 2 & 98 & 898.9117217 & 13 \\
\hline
40 & 94.8178306 & 3 & 99 & 920.4235474 & 10 \\
\hline
41 & 101.5685414 & 2 & 100 & 942.1286420 & 24 \\
\hline
42 & 108.5402790 & 1 & 101 & 964.1753574 & 40 \\
\hline
43 & 115.7702833 & 1 & 102 & 986.4289369 & 56 \\
\hline
44 & 123.1634320 & 1 & 103 & 1008.9408904 & 41 \\
\hline
45 & 130.9053156 & 1 & 104 & 1031.6959327 & 53 \\
\hline
46 & 138.9204719 & 3 & 105 & 1054.8551489 & 58 \\
\hline
47 & 147.1503518 & 5 & 106 & 1078.1682359 & 66 \\
\hline
48 & 155.4174211 & 1 & 107 & 1101.7749883 & 54 \\
\hline
49 & 164.2174643 & 1 & 108 & 1125.5711591 & 57 \\
\hline
50 & 173.0786752 & 1 & 109 & 1149.7450890 & 82 \\
\hline
51 & 182.2666362 & 2 & 110 & 1174.1102793 & 92 \\
\hline
52 & 191.7242791 & 3 & 111 & 1198.7086529 & 60 \\
\hline
53 & 201.3847501 & 2 & 112 & 1223.6244989 & 87 \\
\hline
54 & 211.2834897 & 4 & 113 & 1248.8814345 & 93 \\
\hline
55 & 221.4638138 & 6 & 114 & 1274.3559709 & 118 \\
\hline
56 & 231.8539755 & 3 & 115 & 1300.0999781 & 124 \\
\hline
57 & 242.5180260 & 4 & 116 & 1326.1250674 & 186 \\
\hline
58 & 253.4345983 & 8 & 117 & 1352.3413825 & 232 \\
\hline
59 & 264.5718539 & 4 & 118 & 1378.8765253 & 282 \\
\hline
60 & 275.9094151 & 5 & 119 & 1405.6502758 & 254 \\
\hline
61 & 287.6211392 & 6 & 120 & 1432.6666276 & 208 \\
\hline
62 & 299.4803100 & 2 &  &  &  \\
\hline
\caption{List of global minima for the NE potential. In order, the columns show the number of particles, the NE globally optimal energy, and the number of local minima (including the global minimum).}
\label{table:ne_global}
\end{longtable}
\end{center}

\subsection{Local Minima}
\label{sec:ne_local_minima}

\begin{figure}[h]
	\centering
	\subcaptionbox{\label{fig:ne_minima_bestfit}}{\includegraphics[width=0.45\textwidth]{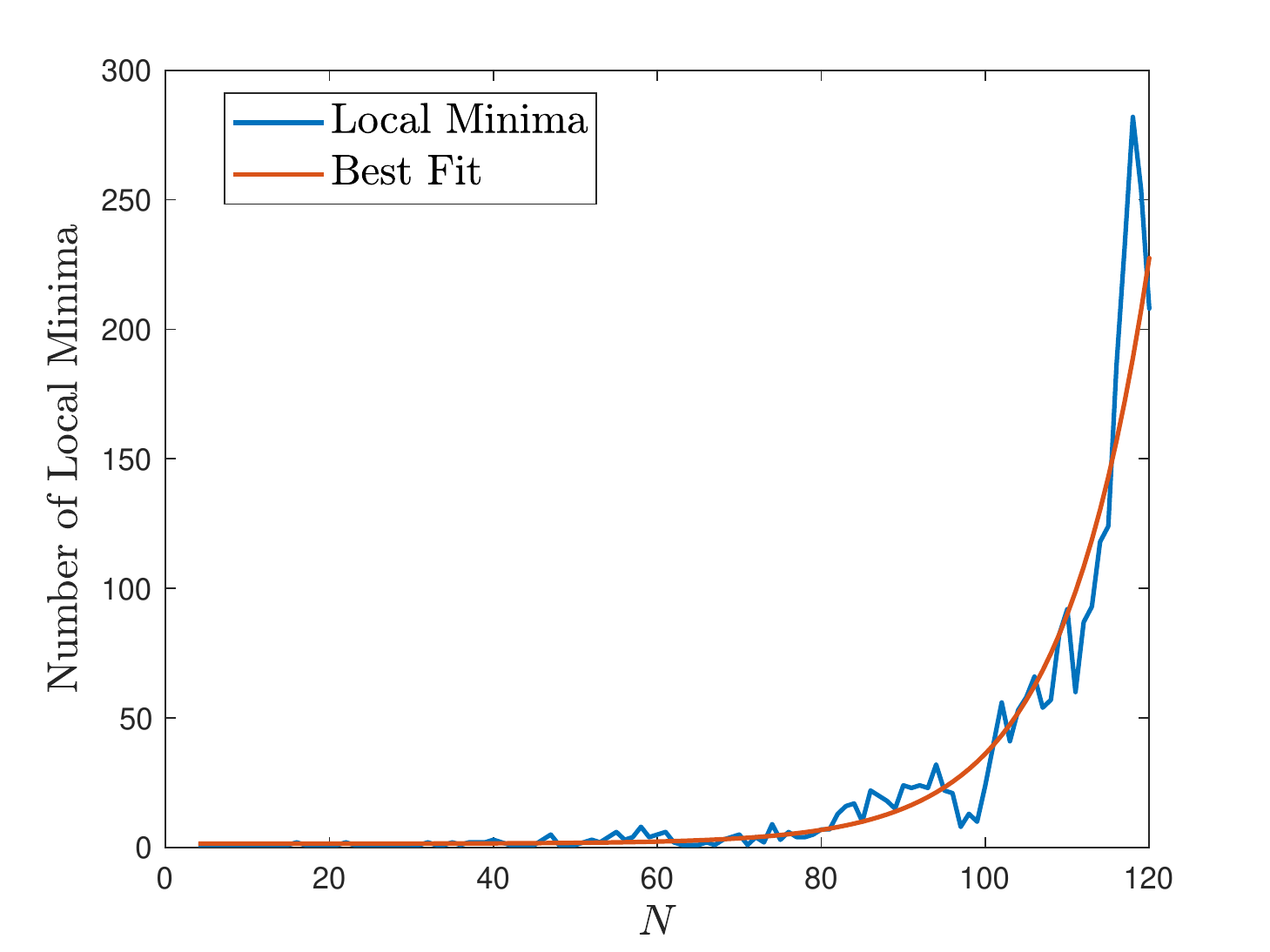}}
	~
	\subcaptionbox{\label{fig:ne_saddles}}{\includegraphics[width=0.45\textwidth]{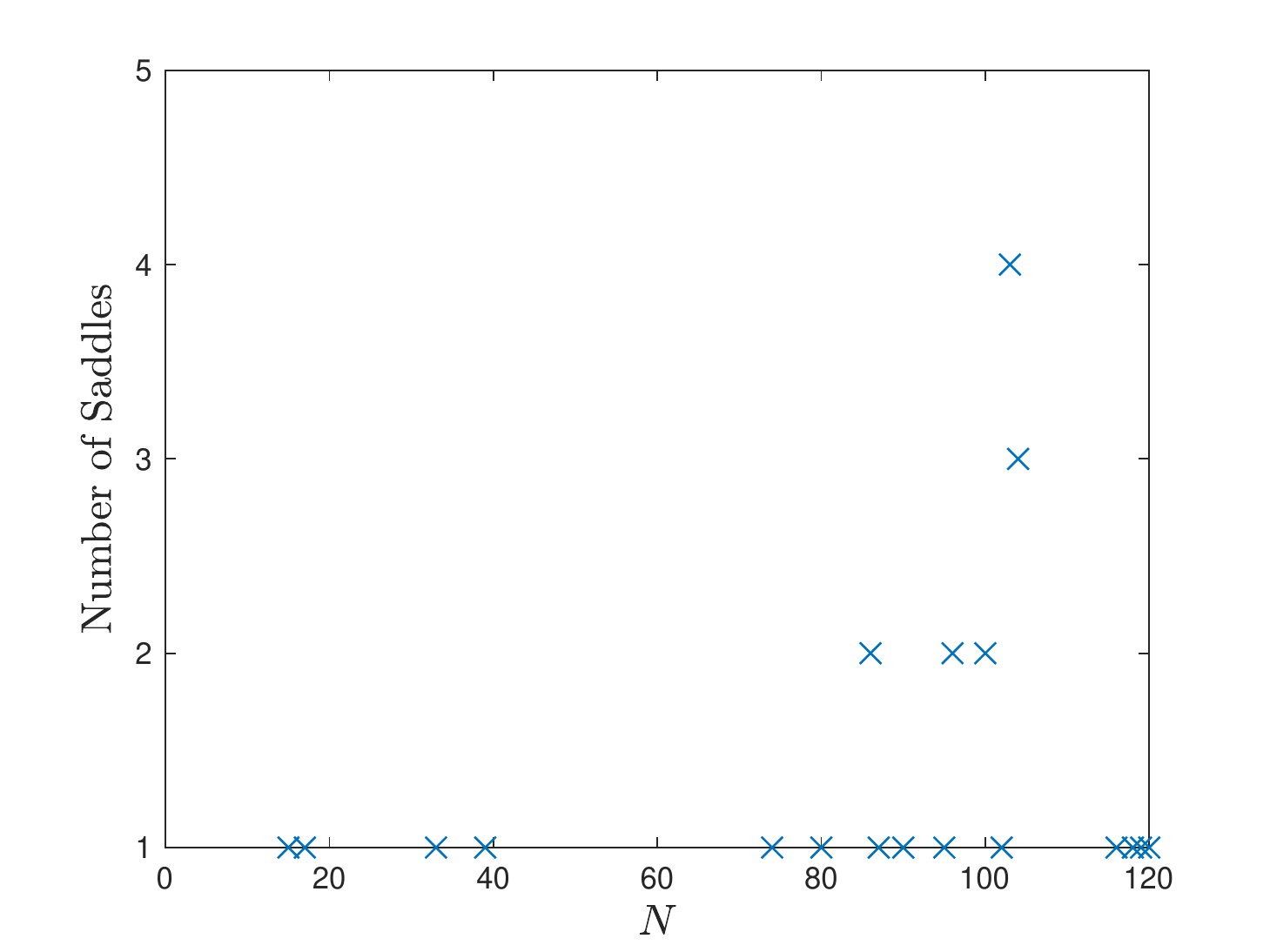}}
	\caption{\subref{fig:ne_minima_bestfit}) Number of minima found for the NE potential and the best-fit curve, Eq. \eqref{eqn:NE_bestfit}. \subref{fig:ne_saddles}) Saddle points found for the NE potential. (Color online).  }
	\label{fig:ne_all_minima}
\end{figure}

The number of computed local minima for the NE potential is shown in Figure \ref{fig:ne_all_minima}.  As the number of minima is expected to grow exponentially, we fit a curve of the form $a_0+a_1e^{a_2N}$ in the least squares sense.  This is a non-linear curve fitting problem which is handled numerically with MATLAB's \emph{lsqcurvefit}() function.  We find
\begin{equation}
	n(N)\approx1.483278 +  0.002963 e^{ 0.093686N},
	\label{eqn:NE_bestfit}
\end{equation}
where $n$ is the best fit number of minima. The reasonable agreement supports the supposed exponential growth rate \cite{mehta2016kinetic}.  

\begin{figure}[h]
	\centering
	\subcaptionbox{\label{fig:spectrum_ne_65_90}}{\includegraphics[width=0.45\textwidth]{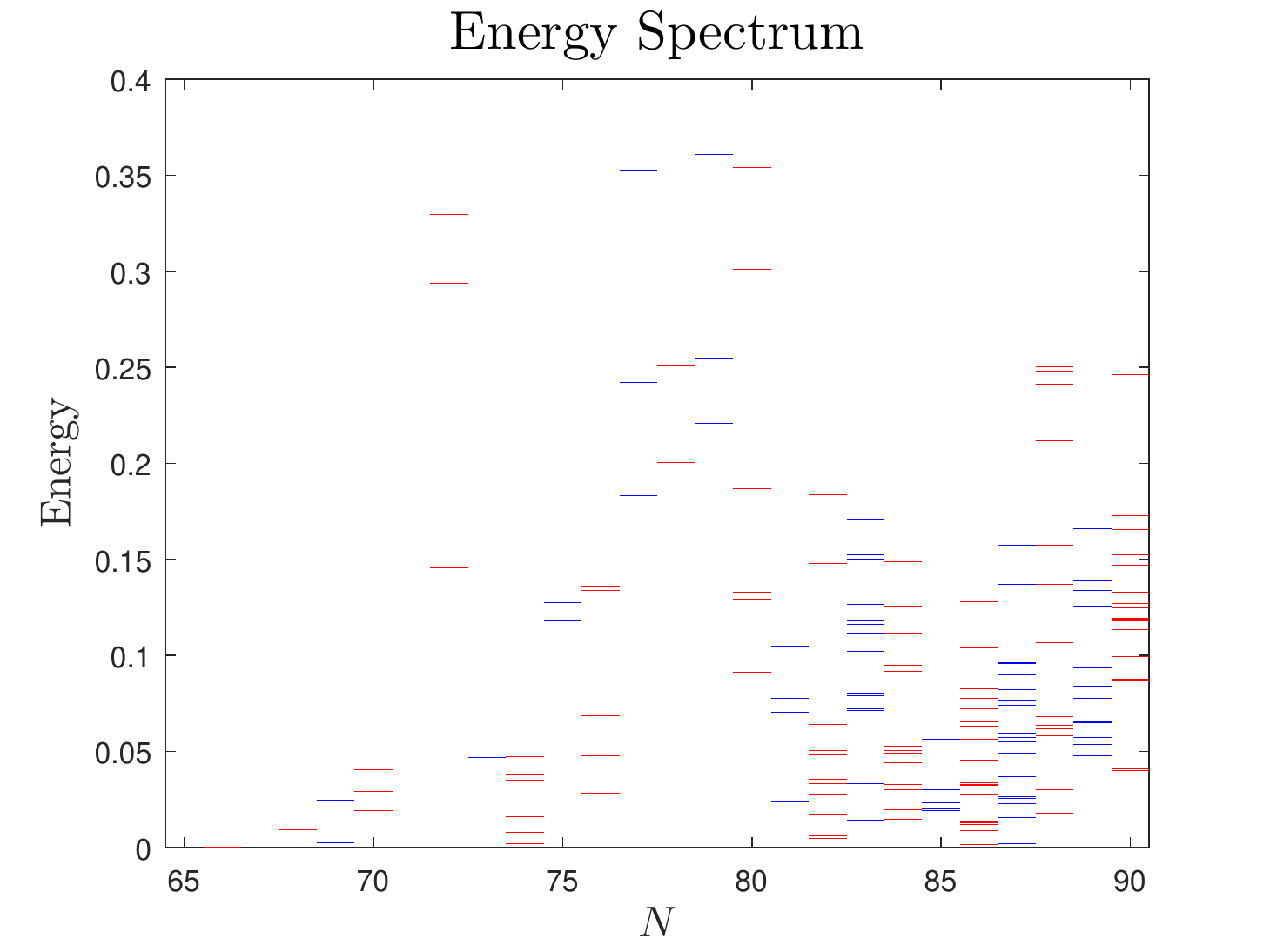}}
	\subcaptionbox{\label{fig:spectrum_ne_90_110}}{\includegraphics[width=0.45\textwidth]{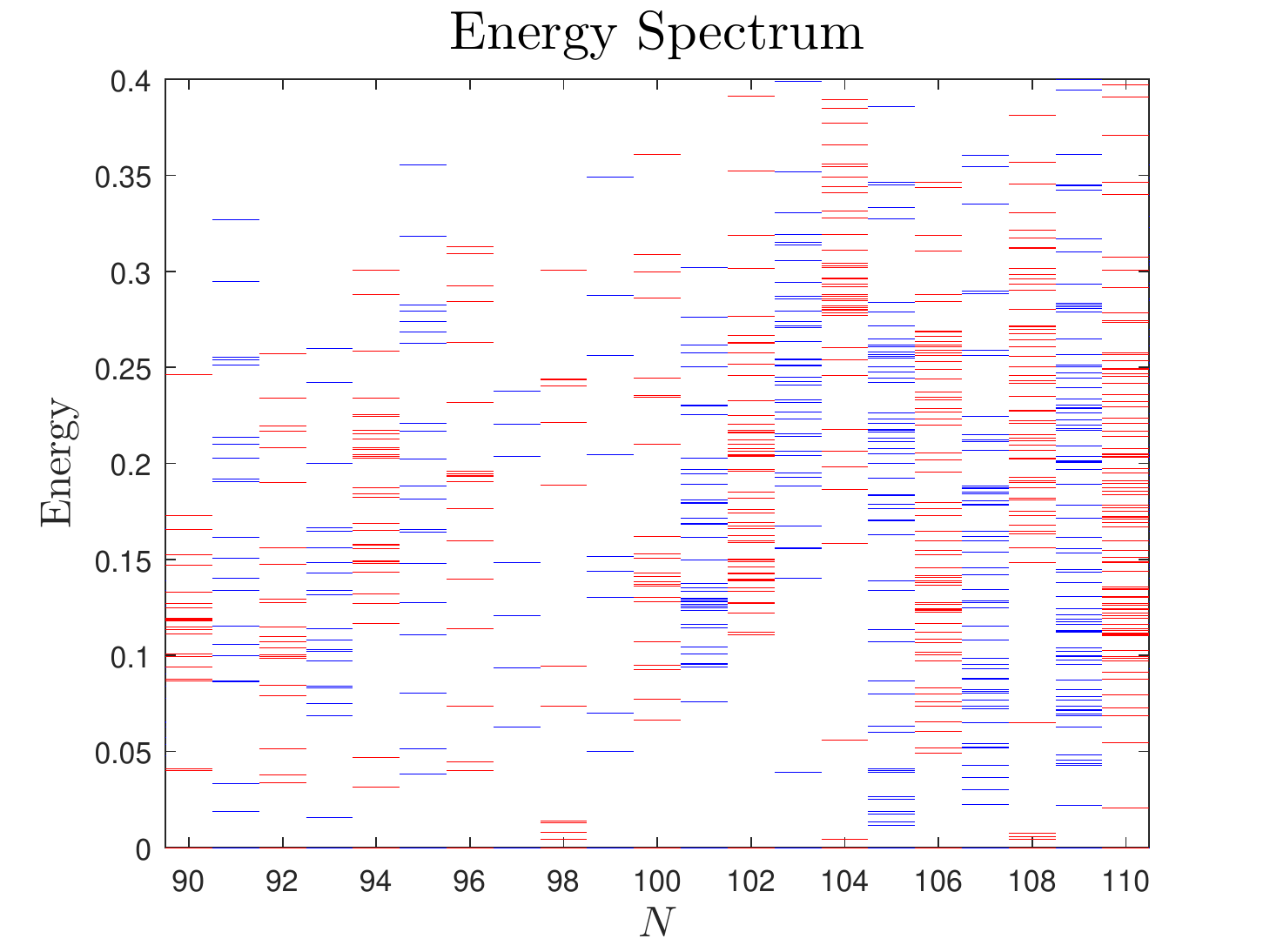}}\\
	\subcaptionbox{\label{fig:spectrum_ne_110_120}}{\includegraphics[width=0.45\textwidth]{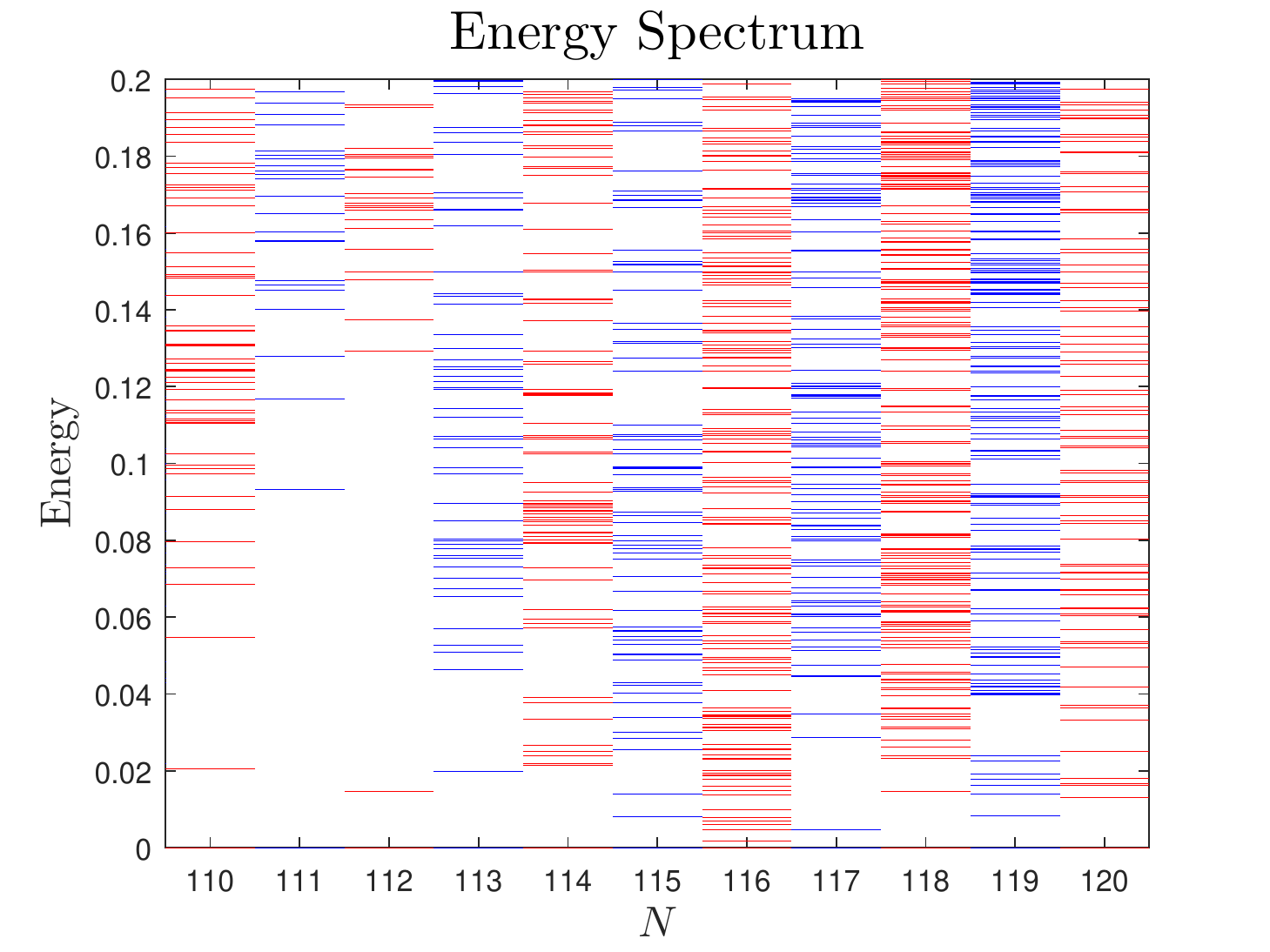}}
	\caption{NE energies of locally optimal configurations relative to the global minima for \subref{fig:spectrum_ne_65_90}) $65\leq N\leq 90$, \subref{fig:spectrum_ne_90_110}) $90\leq N \leq 110$, and \subref{fig:spectrum_ne_110_120}) $110\leq N \leq 120$.  The energies at zero correspond to the global minima.  The vertical axes have been adjusted for clarity.  Note that some minima fall outside the range of the vertical axis. (Color online). }
	\label{fig:spectrum_ne}
\end{figure}

A total of 2780 putative minima and a few saddle points were discovered.  A list of the local and global minima with their corresponding energies and geometric properties are given in Table \ref{table:ne_local} for $N\leq65$. The remaining data for $N\leq120$ are available in a MATLAB file in the online material.  The computed energies are presented as a spectrum in Figure \ref{fig:spectrum_ne}.  They become increasingly dense with $N$.  

\begin{center}
\begin{longtable}{|r|r|r|r|r|r|}
\hline
$N$ & NE Energy & $c_3+c_4+c_5+c_6+c_7$ & $N$ & NE Energy & $c_3+c_4+c_5+c_6+c_7$ \\
\hline
4 & -1.6671799 & 4+0+0+0+0 &  & 147.1833136 & 0+0+12+35+0 \\
\hline
5 & -2.0879876 & 2+3+0+0+0 &  & 147.2448260 & 0+0+12+35+0 \\
\hline
6 & -2.5810055 & 0+6+0+0+0 & 48 & 155.4174211 & 0+0+12+36+0 \\
\hline
7 & -2.7636584 & 0+5+2+0+0 & 49 & 164.2174643 & 0+0+12+37+0 \\
\hline
8 & -2.9495765 & 0+4+4+0+0 & 50 & 173.0786752 & 0+0+12+38+0 \\
\hline
9 & -2.9764336 & 0+3+6+0+0 & 51 & 182.2666362 & 0+0+12+39+0 \\
\hline
10 & -2.8357352 & 0+2+8+0+0 &  & 182.4052019 & 0+0+12+39+0 \\
\hline
11 & -2.4567341 & 0+2+8+1+0 & 52 & 191.7242791 & 0+0+12+40+0 \\
\hline
12 & -2.1612842 & 0+0+12+0+0 &  & 191.7276751 & 0+0+12+40+0 \\
\hline
13 & -1.3678269 & 0+1+10+2+0 &  & 191.7432297 & 0+0+12+40+0 \\
\hline
14 & -0.5525928 & 0+0+12+2+0 & 53 & 201.3847501 & 0+0+12+41+0 \\
\hline
15 & 0.4774376 & 0+0+12+3+0 &  & 201.3952021 & 0+0+12+41+0 \\
\hline
16 & 1.6784049 & 0+0+12+4+0 & 54 & 211.2834897 & 0+0+12+42+0 \\
\hline
 & 1.6888964 & 0+0+12+4+0 &  & 211.2881597 & 0+0+12+42+0 \\
\hline
17 & 3.0751594 & 0+0+12+5+0 &  & 211.2946443 & 0+0+12+42+0 \\
\hline
18 & 4.6651247 & 0+2+8+8+0 &  & 211.2965687 & 0+0+12+42+0 \\
\hline
19 & 6.5461714 & 0+0+12+7+0 & 55 & 221.4638138 & 0+0+12+43+0 \\
\hline
20 & 8.4817896 & 0+0+12+8+0 &  & 221.4688811 & 0+0+14+39+2 \\
\hline
21 & 10.7013196 & 0+1+10+10+0 &  & 221.4690985 & 0+0+12+43+0 \\
\hline
22 & 13.1017418 & 0+0+12+10+0 &  & 221.4816551 & 0+0+12+43+0 \\
\hline
 & 13.1259621 & 0+0+12+10+0 &  & 221.4821066 & 0+0+12+43+0 \\
\hline
23 & 15.8212821 & 0+0+12+11+0 &  & 221.4948761 & 0+0+12+43+0 \\
\hline
24 & 18.5819815 & 0+0+12+12+0 & 56 & 231.8539755 & 0+0+12+44+0 \\
\hline
25 & 21.7249125 & 0+0+12+13+0 &  & 231.8541155 & 0+0+12+44+0 \\
\hline
26 & 25.0100312 & 0+0+12+14+0 &  & 231.8581363 & 0+0+12+44+0 \\
\hline
27 & 28.4296992 & 0+0+12+15+0 & 57 & 242.5180260 & 0+0+12+45+0 \\
\hline
28 & 32.1929330 & 0+0+12+16+0 &  & 242.5606388 & 0+0+13+43+1 \\
\hline
29 & 36.2197826 & 0+0+12+17+0 &  & 242.5722866 & 0+0+12+45+0 \\
\hline
30 & 40.3544394 & 0+0+12+18+0 &  & 242.5742216 & 0+0+12+45+0 \\
\hline
31 & 44.7576167 & 0+0+12+19+0 & 58 & 253.4345983 & 0+0+12+46+0 \\
\hline
32 & 49.2409494 & 0+0+12+20+0 &  & 253.4429647 & 0+0+12+46+0 \\
\hline
 & 49.4893595 & 0+0+12+20+0 &  & 253.4438456 & 0+0+12+46+0 \\
\hline
33 & 54.2959715 & 0+0+13+19+1 &  & 253.4451075 & 0+0+12+46+0 \\
\hline
34 & 59.3794885 & 0+0+12+22+0 &  & 253.4532446 & 0+0+12+46+0 \\
\hline
35 & 64.7367107 & 0+0+12+23+0 &  & 253.4574482 & 0+0+12+46+0 \\
\hline
 & 64.7405651 & 0+0+12+23+0 &  & 253.4675340 & 0+0+12+46+0 \\
\hline
36 & 70.2760966 & 0+0+12+24+0 &  & 253.5691064 & 0+0+14+42+2 \\
\hline
37 & 76.0662374 & 0+0+12+25+0 & 59 & 264.5718539 & 0+0+14+43+2 \\
\hline
 & 76.0768682 & 0+0+12+25+0 &  & 264.5733388 & 0+0+12+47+0 \\
\hline
38 & 82.0802998 & 0+0+12+26+0 &  & 264.5850623 & 0+0+12+47+0 \\
\hline
 & 82.0931587 & 0+0+12+26+0 &  & 264.5916096 & 0+0+12+47+0 \\
\hline
39 & 88.3295602 & 0+0+12+27+0 & 60 & 275.9094151 & 0+0+12+48+0 \\
\hline
 & 88.3900685 & 0+0+12+27+0 &  & 275.9145551 & 0+0+12+48+0 \\
\hline
40 & 94.8178306 & 0+0+12+28+0 &  & 275.9214751 & 0+0+12+48+0 \\
\hline
 & 94.8756103 & 0+0+12+28+0 &  & 276.0688360 & 0+0+12+48+0 \\
\hline
 & 94.8953050 & 0+0+12+28+0 &  & 276.0811919 & 0+0+12+48+0 \\
\hline
41 & 101.5685414 & 0+0+12+29+0 & 61 & 287.6211392 & 0+0+12+49+0 \\
\hline
 & 101.6395277 & 0+0+12+29+0 &  & 287.6322577 & 0+0+12+49+0 \\
\hline
42 & 108.5402790 & 0+0+12+30+0 &  & 287.6359522 & 0+0+12+49+0 \\
\hline
43 & 115.7702833 & 0+0+12+31+0 &  & 287.6534930 & 0+0+12+49+0 \\
\hline
44 & 123.1634320 & 0+0+12+32+0 &  & 287.6561326 & 0+0+12+49+0 \\
\hline
45 & 130.9053156 & 0+0+12+33+0 &  & 287.6655546 & 0+0+12+49+0 \\
\hline
46 & 138.9204719 & 0+0+12+34+0 & 62 & 299.4803100 & 0+0+12+50+0 \\
\hline
 & 138.9242053 & 0+0+12+34+0 &  & 299.5177575 & 0+0+12+50+0 \\
\hline
 & 138.9260526 & 0+0+12+34+0 & 63 & 311.6558511 & 0+0+12+51+0 \\
\hline
47 & 147.1503518 & 0+0+12+35+0 & 64 & 324.0896299 & 0+0+12+52+0 \\
\hline
 & 147.1538854 & 0+0+12+35+0 & 65 & 336.7697094 & 0+0+12+53+0 \\
\hline
 & 147.1659480 & 0+0+12+35+0 &  &  &  \\
\hline
\caption{Comprehensive list of local and global minima for the NE potential (up to $N=65$). In order, the columns show the number of particles, the NE energy, and the numbers of particles with coordination numbers 3 to 7.  The remaining data up to $N=120$ are available online (see the online description). }
\label{table:ne_local}
\end{longtable}
\end{center}

Computation of optimal configurations for the NE potential were also performed by \cite{cheviakov2013narrow}.  We obtain slightly lower energies for $N=90, 95$ and more significantly lower energies for $N=105$ and $N=115$.  The other optimal energies agree with our results up to the given precision.   Many of the energies obtained here are identical to those given in Table 4.2 of \cite{cheviakov2010asymptotic} to the given precision up to $N\approx 35$.  Above $N\approx 35$ the energies given here are slightly lower. 

\begin{figure}[h]
	\centering
	\subcaptionbox{\label{fig:ne_N117_1}}{\includegraphics[width=0.4\textwidth]{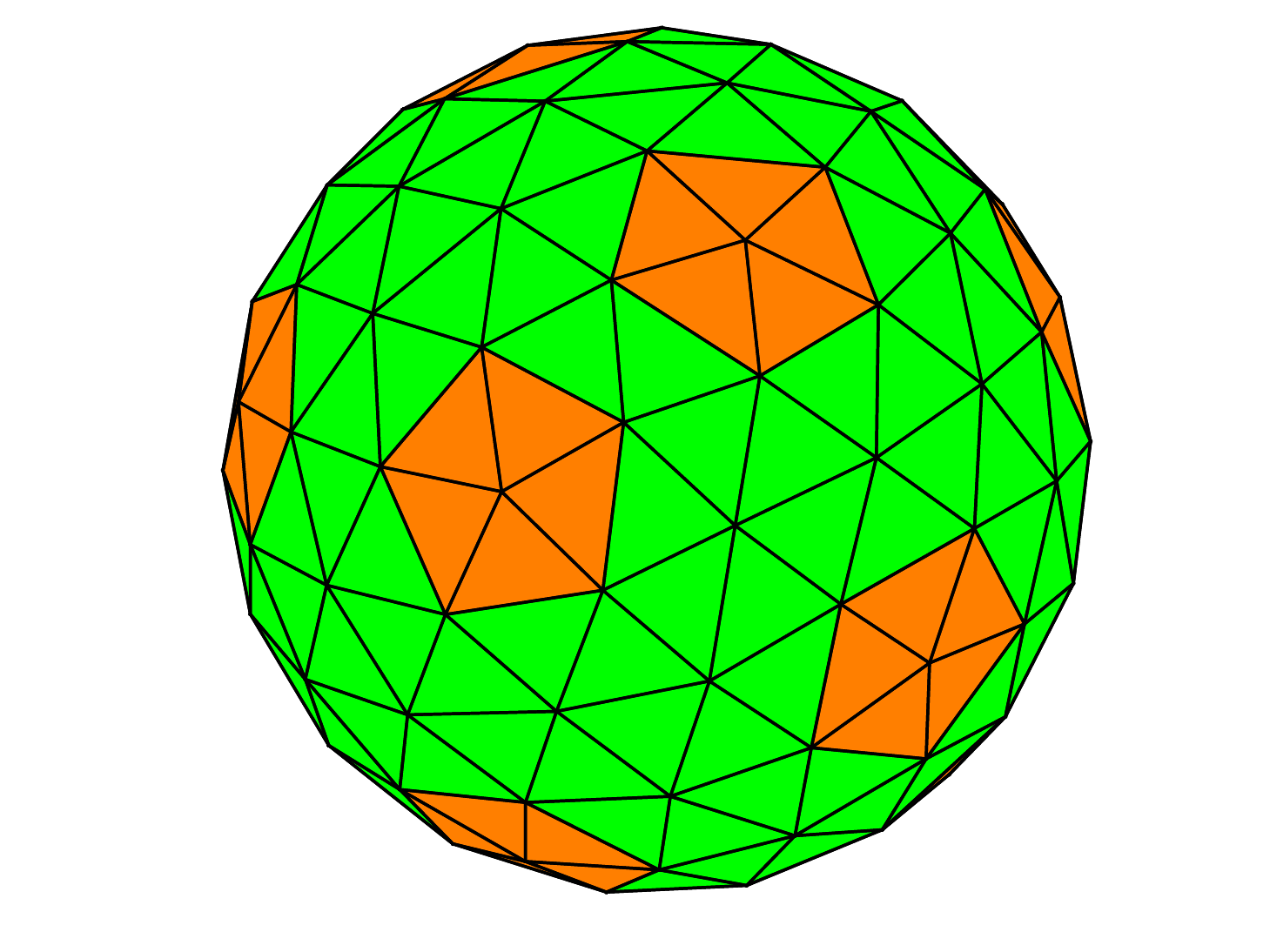}}\hspace{8ex}
	\subcaptionbox{\label{fig:ne_N117_90}}{\includegraphics[width=0.4\textwidth]{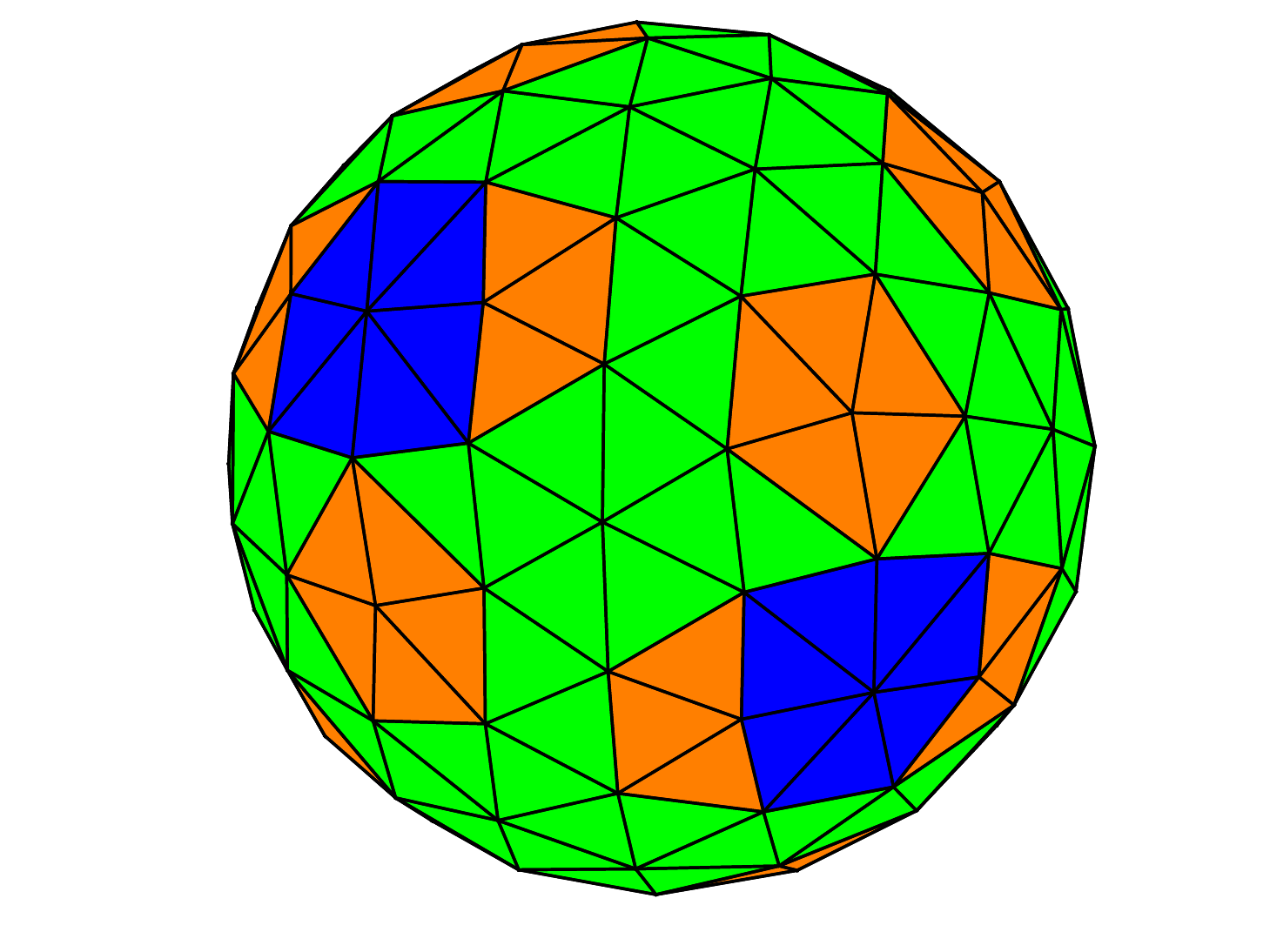}}\\
	\subcaptionbox{\label{fig:ne_N117_91}}{\includegraphics[width=0.4\textwidth]{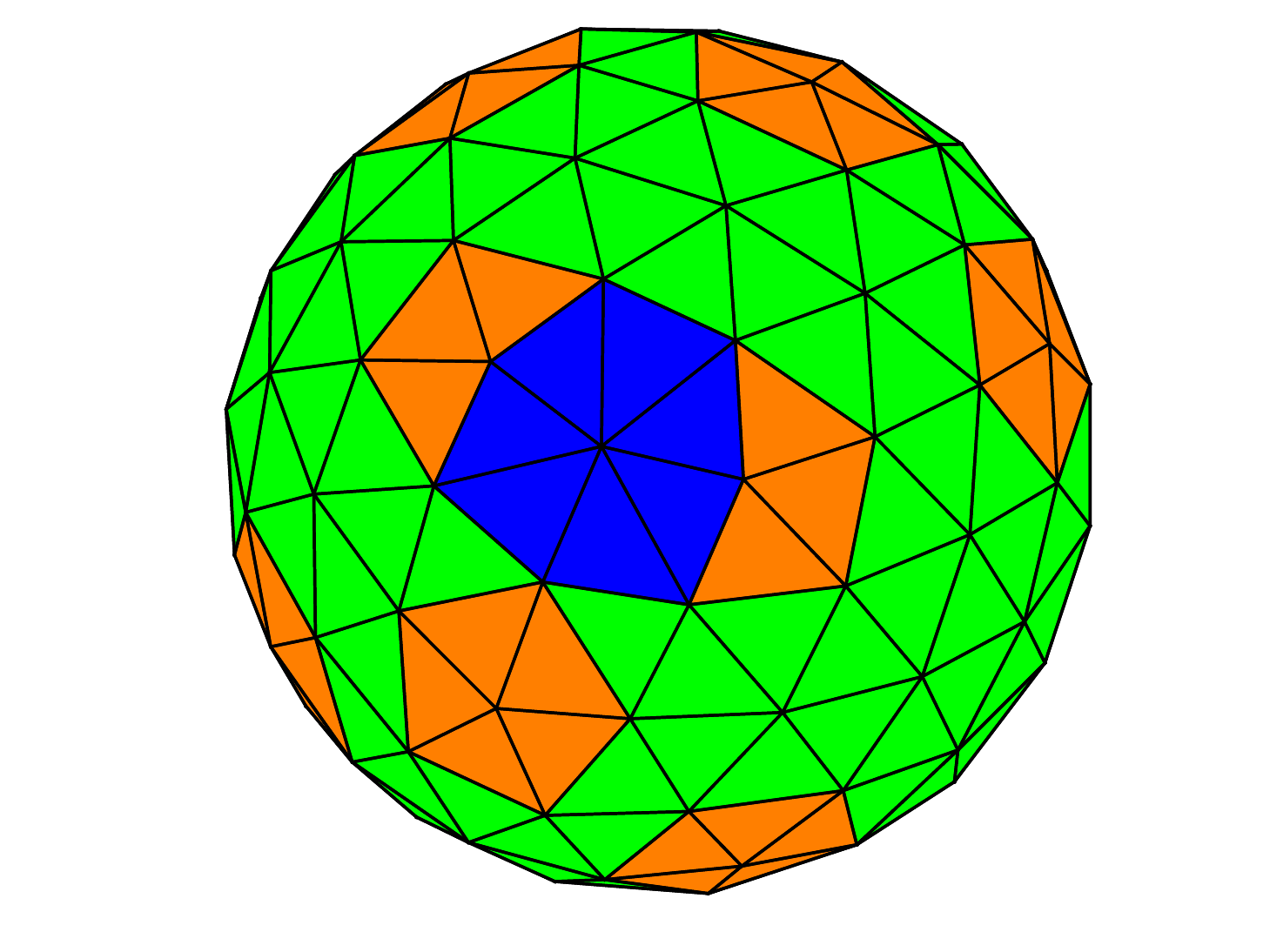}}\\
	\caption{Examples of optimal NE configurations for $N=117$ for which 232 local minima were found. \subref{fig:ne_N117_1}): The global minimum. \subref{fig:ne_N117_90}) \& \subref{fig:ne_N117_91}): Two local minima adjacent in energy.  The computed NE energies are approximately 1352.341, 1352.513, and 1352.514 respectively. (Color online).}
	\label{fig:ne_scars}
\end{figure}

Some examples of computed local minima are given in Figure \ref{fig:ne_scars} showing the scar structure.  In general, there appears to be little or no symmetry in this structure, especially for large $N$.  As a general observation, coordination numbers alone don't provide enough information to distinguish configurations.  This is also seen in columns 3 and 6 of Table \ref{table:ne_local} in which many minima have the same coordination numbers for a given $N$. However configurations with very different scar structures are likely not equivalent geometrically (as in Figure \ref{fig:ne_scars}).

\subsection{A Scaling Law for the Minimal NE Energy}
\label{sec:ne_scaling_law}

In this section, we compare the computed globally optimal energies with an asymptotic scaling law for the pairwise energy \eqref{eqn:ne_energy} valid in the limit $N\rightarrow\infty$.  We rewrite equation \eqref{eqn:ne_energy} as 
\begin{equation}
	\mathcal{H}_\text{NE} = \mathcal{H}_\text{C} + \frac{1}{2}\mathcal{H}_\text{L} + \mathcal{H}_\text{L2}
	\label{eqn:ne_sum}
\end{equation}
where $\mathcal{H}_\text{C}$ and $\mathcal{H}_\text{L}$ are given in equations \eqref{eqn:coul_energy} and \eqref{eqn:log_energy} respectively and 
\begin{equation}
	\mathcal{H}_{\text{L}2}=-\frac{1}{2}\sum_{i<j}^N\log(2+|\mathbf{x}_i-\mathbf{x}_j|).
	\label{eqn:log2_energy}
\end{equation} 

We have the following results from \cite{cheviakov2010asymptotic} (note the factor of $1/2$ difference in our definition of $\mathcal{H}_\text{L}$, equation \eqref{eqn:log_energy}):
\begin{subequations}
\begin{equation}
	\mathcal{H}_\text{C}	\approx \frac{N^2}{2}-\frac{1}{2}N^{3/2}+\frac{1}{12}N^{1/2} + \mathcal{O}(N^{-1/2}),
	\label{eqn:coul_scale_energy}
\end{equation}
\begin{equation}
	\mathcal{H}_\text{L}	\approx \frac{N^2}{4}(1-2\log2)-\frac{N}{4}\log N-\frac{N}{4}(1-2\log2)+\frac{1}{12}\log N-\frac{1}{6}\log2 + \mathcal{O}(N^{-1}),
	\label{eqn:log_scale_energy}
\end{equation}
\begin{equation}
	\mathcal{H}_{\text{L}2}	\approx -\frac{N^2}{8}(1+2\log2) + \frac{\log2}{4}N + \frac{1}{6}N^{1/2} - \left(\frac{1}{16} + \frac{\log2}{12}\right) + \mathcal{O}(N^{-1/2}).
	\label{eqn:log2_scale_energy}
\end{equation}
\end{subequations}
Substituting the above into \eqref{eqn:ne_sum} gives the desired scaling law (also derived in \cite{cheviakov2010asymptotic}):
\begin{IEEEeqnarray}{RCL}
	\mathcal{H}_\text{NE} & \approx & \frac{N^2}{2}\left(1-\log2\right)-\frac{1}{2}N^{3/2}-\frac{N}{8}\log N-\frac{N}{8}\left(1-4\log2\right)  +\frac{1}{4}N^{1/2} +\frac{1}{24}\log N \nonumber\\
	&&-\left(\frac{1}{16}+\frac{\log2}{6}\right) +\mathcal{O}(N^{-1/2}).
	\label{eqn:ne_scale_energy}
\end{IEEEeqnarray} 

\begin{figure}[h]
	\centering
	\includegraphics[width=0.7\textwidth]{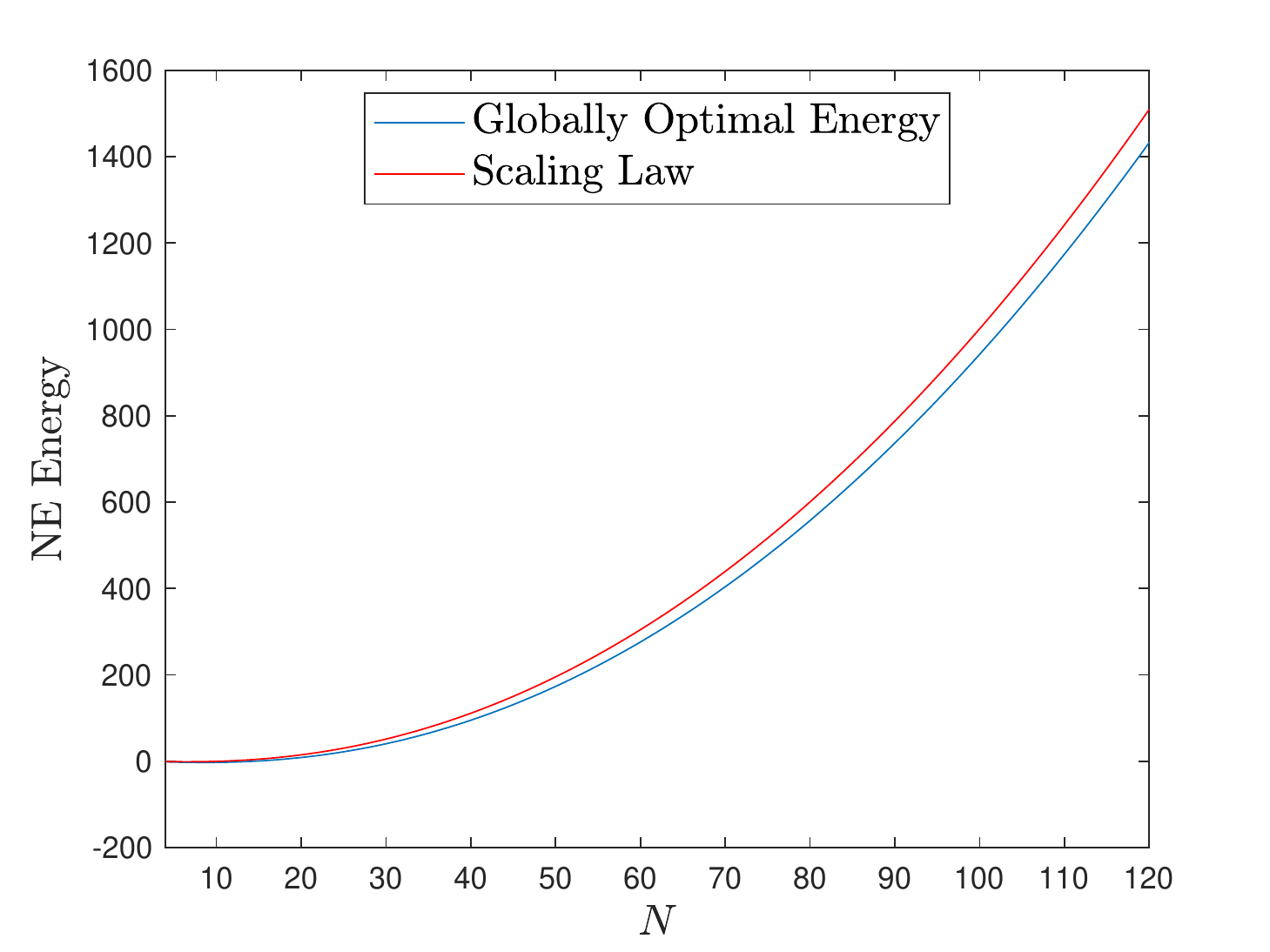}
	\caption{Comparison of globally optimal NE energies with the asymptotic scaling law \eqref{eqn:ne_scale_energy}. (Color online).   }
	\label{fig:ne_scaling}
\end{figure}

Figure \ref{fig:ne_scaling} compares the scaling law \eqref{eqn:ne_scale_energy} with the computed globally optimal energies.  

\section{Results for the NC Potential up to $N = 120$ }
\label{sec:results_NC}

Optimal configurations (both local and global) for the NC potential have not previously been computed to the authors' knowledge.  The current section is organized in the same way as the previous section.  First present tables of the computed global minima and the corresponding energies.  Then we give results for local minima for $N\leq65$ in a table and finally we present a previously derived scaling law \cite{lindsay2017first}.  Data on the local minima for $N> 65$ cannot be included due to the amount of data and are available online (see the online description).

\subsection{Global Minima}

Table \ref{table:nc_global} gives the computed globally optimal energies for each $N$ along with the number of computed local minima. As with the NE potential, the computation time grows rapidly with $N$.  Between $N=115$ and $N=120$ the number of starting configurations grows from approximately 27000 to 62000, roughly the same as the NE potential.  

\begin{center}
\begin{longtable}{|r|r|r|r|r|r|}
\hline
$N$ & Global NC Energy & Local Minima & $N$ & Global NC Energy & Local Minima \\
\hline
4 & 1.2753079 & 1 & 63 & 761.7365139 & 1 \\
\hline
5 & 2.3325195 & 1 & 64 & 787.7433643 & 1 \\
\hline
6 & 3.6573191 & 1 & 65 & 814.1954515 & 1 \\
\hline
7 & 5.4188195 & 1 & 66 & 841.0627048 & 2 \\
\hline
8 & 7.4781865 & 1 & 67 & 868.3435424 & 1 \\
\hline
9 & 9.9113144 & 1 & 68 & 896.1835478 & 3 \\
\hline
10 & 12.7273233 & 1 & 69 & 924.4370909 & 4 \\
\hline
11 & 15.9633917 & 1 & 70 & 953.1053374 & 5 \\
\hline
12 & 19.4448610 & 1 & 71 & 982.2534837 & 2 \\
\hline
13 & 23.4988136 & 1 & 72 & 1011.7220316 & 5 \\
\hline
14 & 27.8546601 & 1 & 73 & 1041.9378711 & 3 \\
\hline
15 & 32.6249857 & 1 & 74 & 1072.4773046 & 9 \\
\hline
16 & 37.7841349 & 2 & 75 & 1103.3861677 & 3 \\
\hline
17 & 43.3480585 & 1 & 76 & 1134.7989611 & 6 \\
\hline
18 & 49.3152900 & 1 & 77 & 1166.5921261 & 4 \\
\hline
19 & 55.7454616 & 1 & 78 & 1198.9012840 & 4 \\
\hline
20 & 62.4925763 & 1 & 79 & 1231.7114783 & 5 \\
\hline
21 & 69.7021610 & 1 & 80 & 1264.9343750 & 8 \\
\hline
22 & 77.3074771 & 2 & 81 & 1298.7071786 & 7 \\
\hline
23 & 85.3990644 & 1 & 82 & 1332.8686991 & 13 \\
\hline
24 & 93.7956788 & 1 & 83 & 1367.4479328 & 16 \\
\hline
25 & 102.7220631 & 1 & 84 & 1402.5044669 & 17 \\
\hline
26 & 112.0190373 & 1 & 85 & 1438.0212839 & 10 \\
\hline
27 & 121.6815612 & 1 & 86 & 1473.9935322 & 23 \\
\hline
28 & 131.8513551 & 1 & 87 & 1510.4080648 & 21 \\
\hline
29 & 142.4737389 & 1 & 88 & 1547.2565862 & 19 \\
\hline
30 & 153.4431719 & 1 & 89 & 1584.6017630 & 19 \\
\hline
31 & 164.8676908 & 1 & 90 & 1622.3838285 & 26 \\
\hline
32 & 176.6198170 & 2 & 91 & 1660.6362010 & 24 \\
\hline
33 & 189.0434647 & 1 & 92 & 1699.3452879 & 28 \\
\hline
34 & 201.7545872 & 1 & 93 & 1738.5291591 & 27 \\
\hline
35 & 214.9282924 & 2 & 94 & 1778.1304687 & 39 \\
\hline
36 & 228.4996702 & 1 & 95 & 1818.2268735 & 24 \\
\hline
37 & 242.5158012 & 2 & 96 & 1858.7481974 & 23 \\
\hline
38 & 256.9600891 & 2 & 97 & 1899.7641064 & 11 \\
\hline
39 & 271.8381200 & 2 & 98 & 1941.1954839 & 18 \\
\hline
40 & 287.1550632 & 3 & 99 & 1983.1491796 & 14 \\
\hline
41 & 302.9272985 & 2 & 100 & 2025.5050024 & 27 \\
\hline
42 & 319.1244788 & 1 & 101 & 2068.3830362 & 41 \\
\hline
43 & 335.7734900 & 1 & 102 & 2111.6744155 & 60 \\
\hline
44 & 352.8052329 & 1 & 103 & 2155.4211861 & 50 \\
\hline
45 & 370.3587315 & 1 & 104 & 2199.6110343 & 61 \\
\hline
46 & 388.3757521 & 3 & 105 & 2244.3738904 & 63 \\
\hline
47 & 406.8113090 & 5 & 106 & 2289.5077641 & 73 \\
\hline
48 & 425.5351845 & 1 & 107 & 2335.1252008 & 60 \\
\hline
49 & 444.9188072 & 1 & 108 & 2381.1411090 & 65 \\
\hline
50 & 464.6065564 & 1 & 109 & 2427.7106073 & 94 \\
\hline
51 & 484.7997393 & 2 & 110 & 2474.6800384 & 101 \\
\hline
52 & 505.4557374 & 4 & 111 & 2522.0823797 & 75 \\
\hline
53 & 526.5221776 & 2 & 112 & 2569.9899275 & 107 \\
\hline
54 & 548.0282546 & 4 & 113 & 2618.4209072 & 111 \\
\hline
55 & 570.0046269 & 6 & 114 & 2667.2735190 & 140 \\
\hline
56 & 592.3987462 & 3 & 115 & 2716.5898475 & 159 \\
\hline
57 & 615.2584023 & 5 & 116 & 2766.3811001 & 215 \\
\hline
58 & 638.5661060 & 8 & 117 & 2816.5721460 & 265 \\
\hline
59 & 662.2992461 & 4 & 118 & 2867.2673880 & 320 \\
\hline
60 & 686.4417704 & 5 & 119 & 2918.4018751 & 284 \\
\hline
61 & 711.1280295 & 6 & 120 & 2969.9778301 & 238 \\
\hline
62 & 736.1836003 & 3 &  &  &  \\
\hline
\caption{List of global minima for the NC potential. In order, the columns show the number of particles, the NC globally optimal energy, and the number of local minima.}
\label{table:nc_global}
\end{longtable}
\end{center}

\subsection{Local Minima}

A total $3170$ of putative minima and a few saddle points were found.  A list of minima and corresponding geometric properties for $N\leq65$ is given in Table \ref{table:nc_local}.  A complete list up to $N=120$ is available online in a MATLAB file.  

\begin{figure}[h]
	\centering
	\subcaptionbox{\label{fig:nc_minima_bestfit}}{\includegraphics[width=0.45\textwidth]{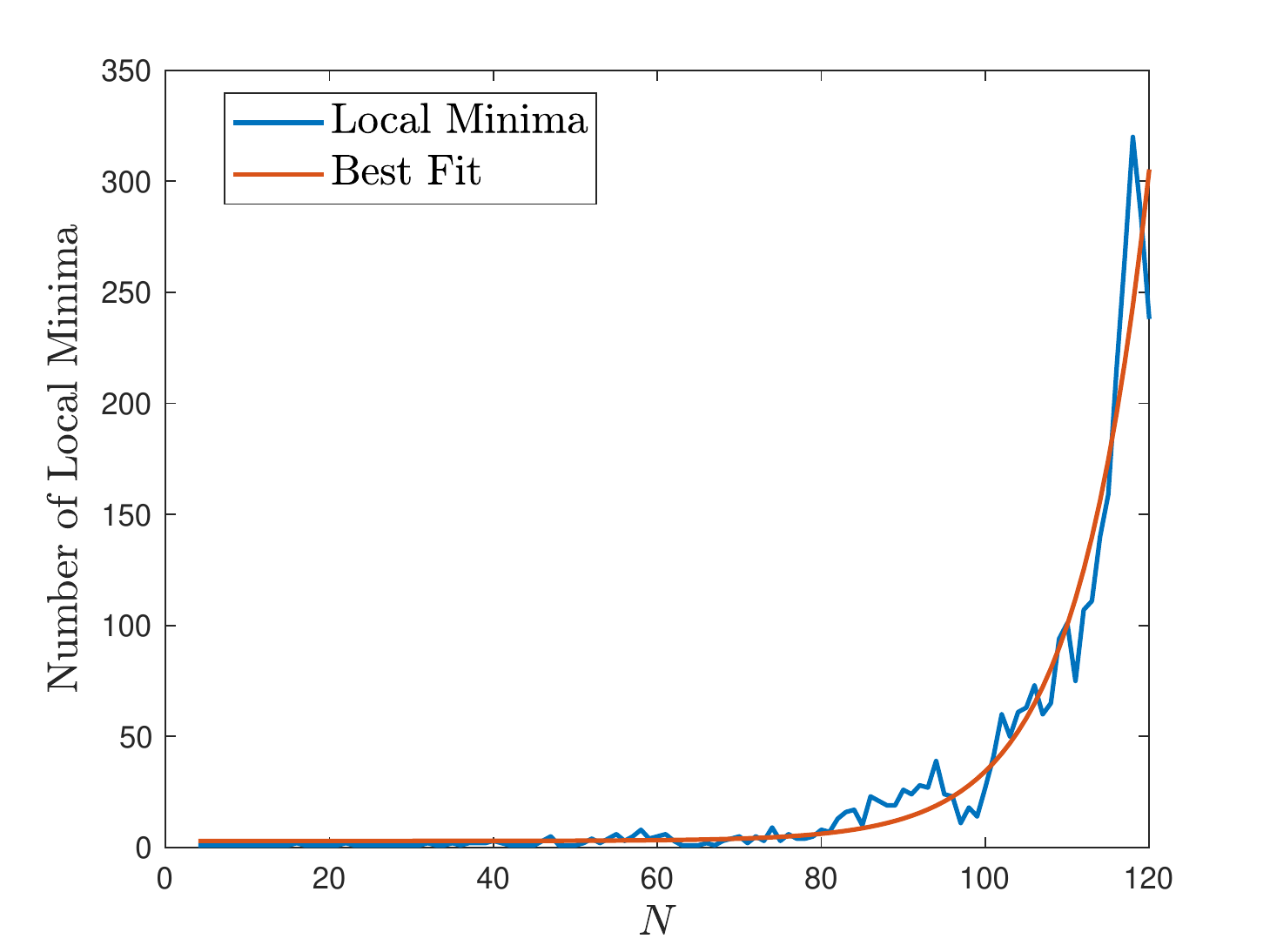}}
	~
	\subcaptionbox{\label{fig:nc_saddles}}{\includegraphics[width=0.45\textwidth]{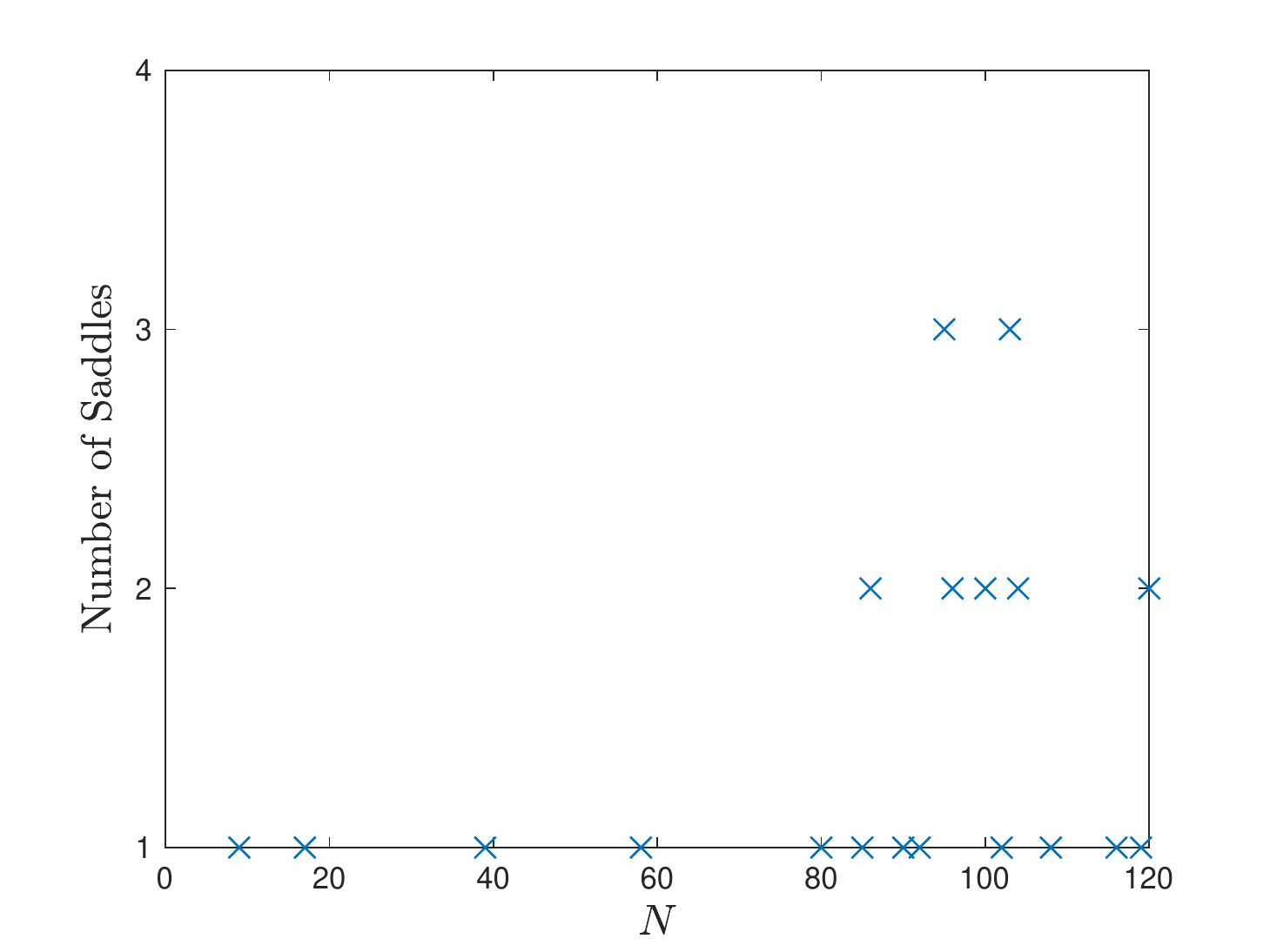}}
	\caption{\subref{fig:nc_minima_bestfit}) Number of minima found for the NC potential and the best-fit curve, Eq. \eqref{eqn:NC_bestfit}. \subref{fig:nc_saddles}) Saddle points found for the NC potential.  (Color online).}
	\label{fig:nc_all_minima}
\end{figure}

As with the NE potential, we fit a curve of the form $n(N)=a_0+a_1e^{a_2N}$ to the number of local minima, where $n$ is the best fit number of minima.  Using the same procedure as described in Section \ref{sec:ne_local_minima}, we find 
\begin{equation}
	n(N)\approx2.968630 +  0.000376e^{0.113313N}.
	\label{eqn:NC_bestfit}
\end{equation}
Figure \ref{fig:nc_all_minima} shows the number of optimal configurations before and after removal of saddle points along with the best fit curve \eqref{eqn:NC_bestfit}.

\begin{figure}[h]
	\centering
	\subcaptionbox{\label{fig:spectrum_nc_65_90}}{\includegraphics[width=0.45\textwidth]{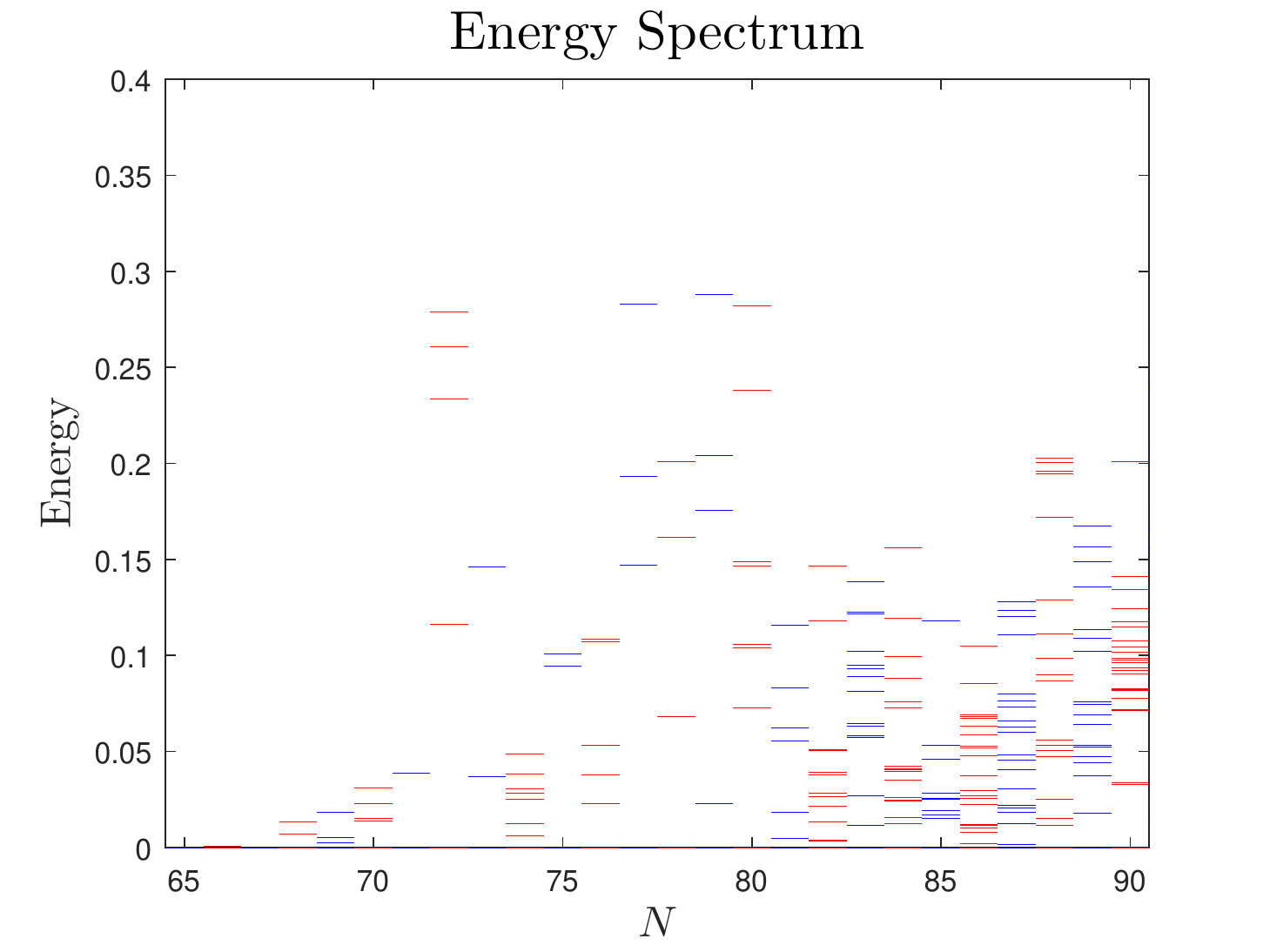}}
	\subcaptionbox{\label{fig:spectrum_nc_90_110}}{\includegraphics[width=0.45\textwidth]{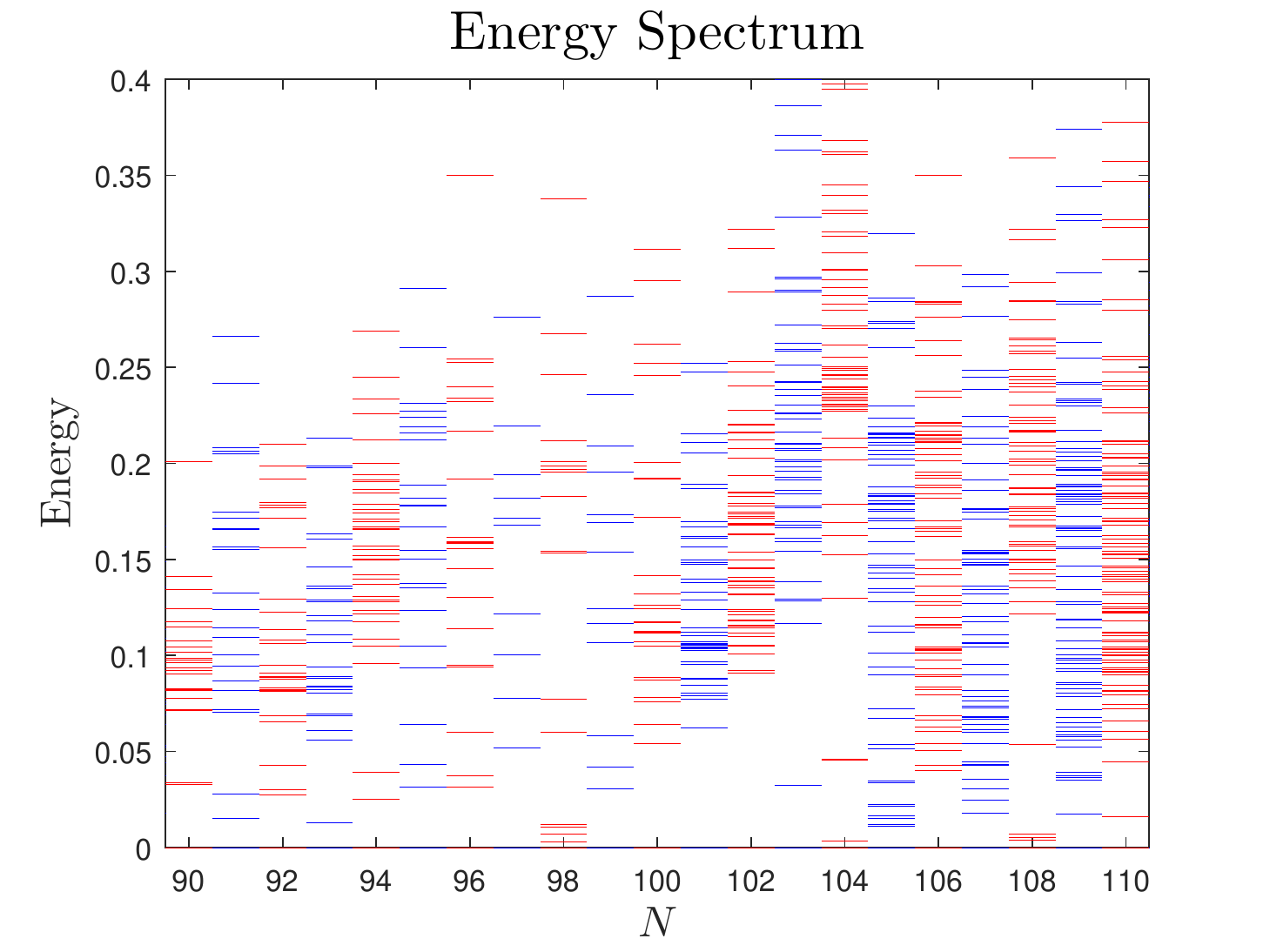}}\\
	\subcaptionbox{\label{fig:spectrum_nc_110_120}}{\includegraphics[width=0.45\textwidth]{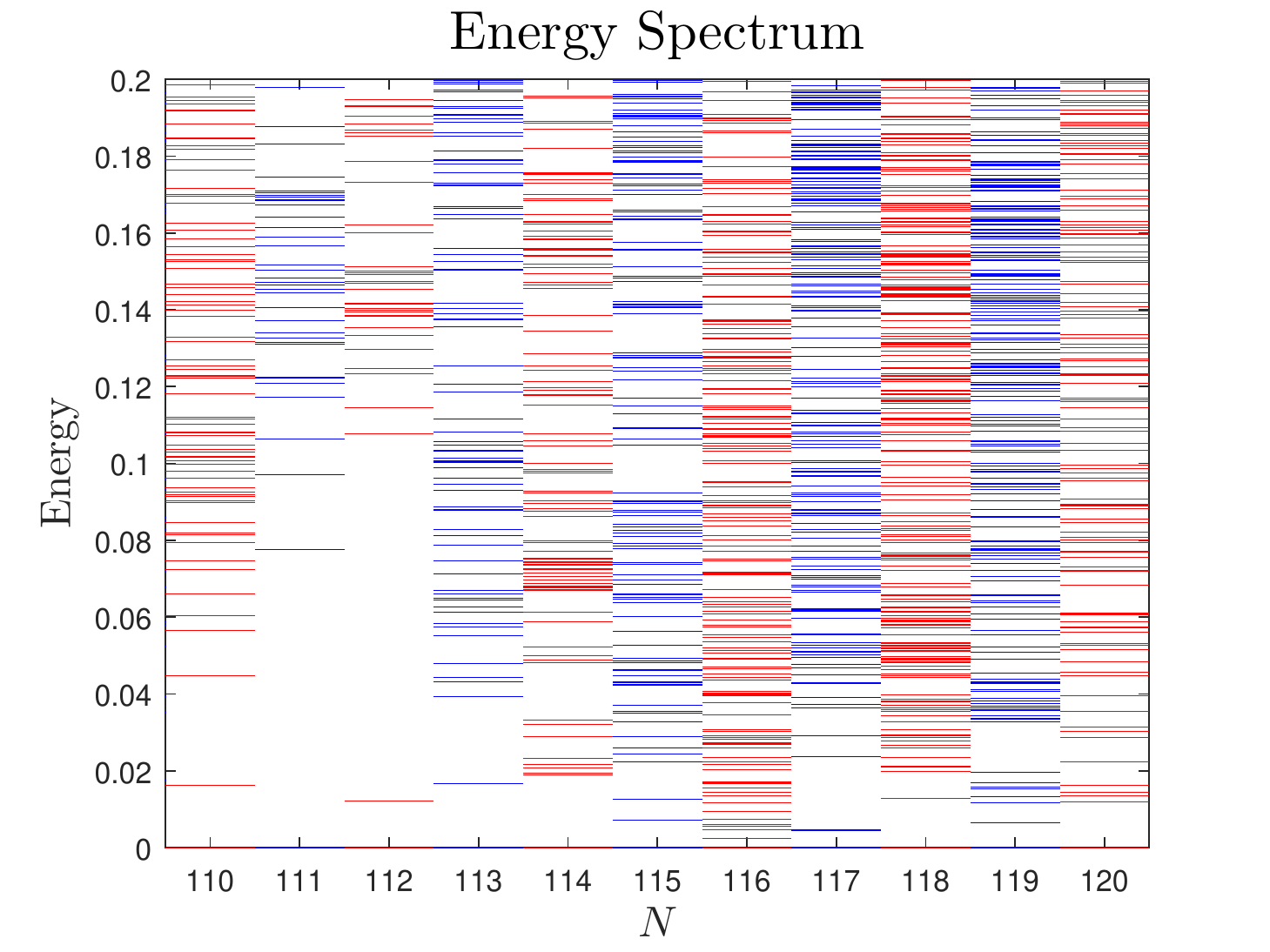}}
	\caption{NC energies of locally optimal configurations relative to the global minima for \subref{fig:spectrum_nc_65_90}) $65\leq N\leq 90$, \subref{fig:spectrum_nc_90_110}) $90\leq N \leq 110$, and \subref{fig:spectrum_nc_110_120}) $110\leq N \leq 120$.  The energies at zero correspond to the global minima.  The vertical axes have been adjusted for clarity.  Note that some minima fall outside the range of the vertical axis. (Color online). }
	\label{fig:spectrum_nc}
\end{figure}

 A spectrum plot of the locally optimal energies is given in Figure \ref{fig:spectrum_nc}. The minima become increasingly dense as $N$ increases.

\begin{center}
\begin{longtable}{|r|r|r|r|r|r|}
\hline
$N$ & NC Energy & $c_3+c_4+c_5+c_6+c_7$ & $N$ & NC Energy & $c_3+c_4+c_5+c_6+c_7$ \\
\hline
4 & 1.2753079 & 4+0+0+0+0 &  & 406.8826917 & 0+0+12+35+0 \\
\hline
5 & 2.3325195 & 2+3+0+0+0 & 48 & 425.5351845 & 0+0+12+36+0 \\
\hline
6 & 3.6573191 & 0+6+0+0+0 & 49 & 444.9188072 & 0+0+12+37+0 \\
\hline
7 & 5.4188195 & 0+5+2+0+0 & 50 & 464.6065564 & 0+0+12+38+0 \\
\hline
8 & 7.4781865 & 0+4+4+0+0 & 51 & 484.7997393 & 0+0+12+39+0 \\
\hline
9 & 9.9113144 & 0+3+6+0+0 &  & 484.9059577 & 0+0+12+39+0 \\
\hline
10 & 12.7273233 & 0+2+8+0+0 & 52 & 505.4557374 & 0+0+12+40+0 \\
\hline
11 & 15.9633917 & 0+2+8+1+0 &  & 505.4584264 & 0+0+12+40+0 \\
\hline
12 & 19.4448610 & 0+0+12+0+0 &  & 505.4705912 & 0+0+12+40+0 \\
\hline
13 & 23.4988136 & 0+1+10+2+0 &  & 505.4723820 & 0+0+12+40+0 \\
\hline
14 & 27.8546601 & 0+0+12+2+0 & 53 & 526.5221776 & 0+0+12+41+0 \\
\hline
15 & 32.6249857 & 0+0+12+3+0 &  & 526.5304615 & 0+0+12+41+0 \\
\hline
16 & 37.7841349 & 0+0+12+4+0 & 54 & 548.0282546 & 0+0+12+42+0 \\
\hline
 & 37.7913424 & 0+0+12+4+0 &  & 548.0313595 & 0+0+12+42+0 \\
\hline
17 & 43.3480585 & 0+0+12+5+0 &  & 548.0367702 & 0+0+12+42+0 \\
\hline
18 & 49.3152900 & 0+2+8+8+0 &  & 548.0387819 & 0+0+12+42+0 \\
\hline
19 & 55.7454616 & 0+0+12+7+0 & 55 & 570.0046269 & 0+0+12+43+0 \\
\hline
20 & 62.4925763 & 0+0+12+8+0 &  & 570.0081782 & 0+0+12+43+0 \\
\hline
21 & 69.7021610 & 0+1+10+10+0 &  & 570.0086879 & 0+0+14+39+2 \\
\hline
22 & 77.3074771 & 0+0+12+10+0 &  & 570.0189717 & 0+0+12+43+0 \\
\hline
 & 77.3248232 & 0+0+12+10+0 &  & 570.0192368 & 0+0+12+43+0 \\
\hline
23 & 85.3990644 & 0+0+12+11+0 &  & 570.0289506 & 0+0+12+43+0 \\
\hline
24 & 93.7956788 & 0+0+12+12+0 & 56 & 592.3987462 & 0+0+12+44+0 \\
\hline
25 & 102.7220631 & 0+0+12+13+0 &  & 592.3992926 & 0+0+12+44+0 \\
\hline
26 & 112.0190373 & 0+0+12+14+0 &  & 592.4021497 & 0+0+12+44+0 \\
\hline
27 & 121.6815612 & 0+0+12+15+0 & 57 & 615.2584023 & 0+0+12+45+0 \\
\hline
28 & 131.8513551 & 0+0+12+16+0 &  & 615.2702496 & 0+0+12+45+0 \\
\hline
29 & 142.4737389 & 0+0+12+17+0 &  & 615.2905050 & 0+0+13+43+1 \\
\hline
30 & 153.4431719 & 0+0+12+18+0 &  & 615.2995971 & 0+0+12+45+0 \\
\hline
31 & 164.8676908 & 0+0+12+19+0 &  & 615.2996416 & 0+0+12+45+0 \\
\hline
32 & 176.6198170 & 0+0+12+20+0 & 58 & 638.5661060 & 0+0+12+46+0 \\
\hline
 & 176.7944252 & 0+0+12+20+0 &  & 638.5724213 & 0+0+12+46+0 \\
\hline
33 & 189.0434647 & 0+0+13+19+1 &  & 638.5731306 & 0+0+12+46+0 \\
\hline
34 & 201.7545872 & 0+0+12+22+0 &  & 638.5737991 & 0+0+12+46+0 \\
\hline
35 & 214.9282924 & 0+0+12+23+0 &  & 638.5790228 & 0+0+12+46+0 \\
\hline
 & 214.9315690 & 0+0+12+23+0 &  & 638.5834576 & 0+0+12+46+0 \\
\hline
36 & 228.4996702 & 0+0+12+24+0 &  & 638.5846113 & 0+0+12+46+0 \\
\hline
37 & 242.5158012 & 0+0+12+25+0 &  & 638.5919380 & 0+0+12+46+0 \\
\hline
 & 242.5235820 & 0+0+12+25+0 & 59 & 662.2992461 & 0+0+14+43+2 \\
\hline
38 & 256.9600891 & 0+0+12+26+0 &  & 662.2999789 & 0+0+12+47+0 \\
\hline
 & 256.9687260 & 0+0+12+26+0 &  & 662.3091948 & 0+0+12+47+0 \\
\hline
39 & 271.8381200 & 0+0+12+27+0 &  & 662.3147258 & 0+0+12+47+0 \\
\hline
 & 271.8830066 & 0+0+12+27+0 & 60 & 686.4417704 & 0+0+12+48+0 \\
\hline
40 & 287.1550632 & 0+0+12+28+0 &  & 686.4460598 & 0+0+12+48+0 \\
\hline
 & 287.1986056 & 0+0+12+28+0 &  & 686.4520078 & 0+0+12+48+0 \\
\hline
 & 287.2116587 & 0+0+12+28+0 &  & 686.5639093 & 0+0+12+48+0 \\
\hline
41 & 302.9272985 & 0+0+12+29+0 &  & 686.5741767 & 0+0+12+48+0 \\
\hline
 & 302.9816133 & 0+0+12+29+0 & 61 & 711.1280295 & 0+0+12+49+0 \\
\hline
42 & 319.1244788 & 0+0+12+30+0 &  & 711.1365911 & 0+0+12+49+0 \\
\hline
43 & 335.7734900 & 0+0+12+31+0 &  & 711.1399905 & 0+0+12+49+0 \\
\hline
44 & 352.8052329 & 0+0+12+32+0 &  & 711.1533041 & 0+0+12+49+0 \\
\hline
45 & 370.3587315 & 0+0+12+33+0 &  & 711.1553205 & 0+0+12+49+0 \\
\hline
46 & 388.3757521 & 0+0+12+34+0 &  & 711.1624250 & 0+0+12+49+0 \\
\hline
 & 388.3783317 & 0+0+12+34+0 & 62 & 736.1836003 & 0+0+12+50+0 \\
\hline
 & 388.3791178 & 0+0+12+34+0 &  & 736.2003615 & 0+0+12+50+0 \\
\hline
47 & 406.8113090 & 0+0+12+35+0 &  & 736.2121671 & 0+0+12+50+0 \\
\hline
 & 406.8138763 & 0+0+12+35+0 & 63 & 761.7365139 & 0+0+12+51+0 \\
\hline
 & 406.8222118 & 0+0+12+35+0 & 64 & 787.7433643 & 0+0+12+52+0 \\
\hline
 & 406.8368865 & 0+0+12+35+0 & 65 & 814.1954515 & 0+0+12+53+0 \\
\hline
\caption{Comprehensive list of local and global minima for the NC potential up to $N=65$. In order, the columns show the number of particles, the NC energy, and the numbers of particles with coordination numbers 3 to 7. The remaining data up to $N=120$ are available online (see the online description).}
\label{table:nc_local}
\end{longtable}
\end{center}

\begin{figure}[h]
	\centering
	\subcaptionbox{\label{fig:nc_N120_1}}{\includegraphics[width=0.4\textwidth]{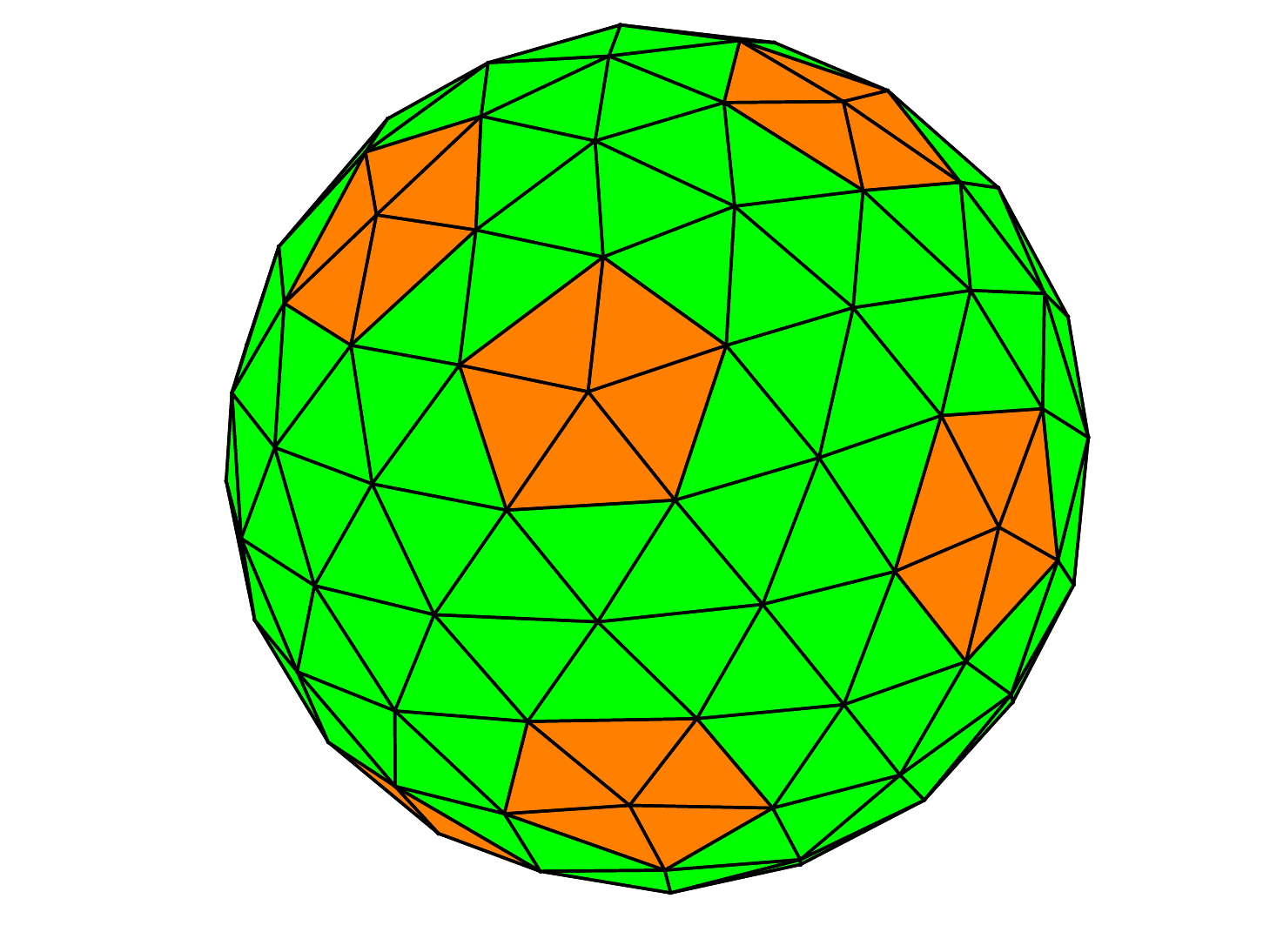}}\hspace{8ex}
	\subcaptionbox{\label{fig:nc_N120_223}}{\includegraphics[width=0.4\textwidth]{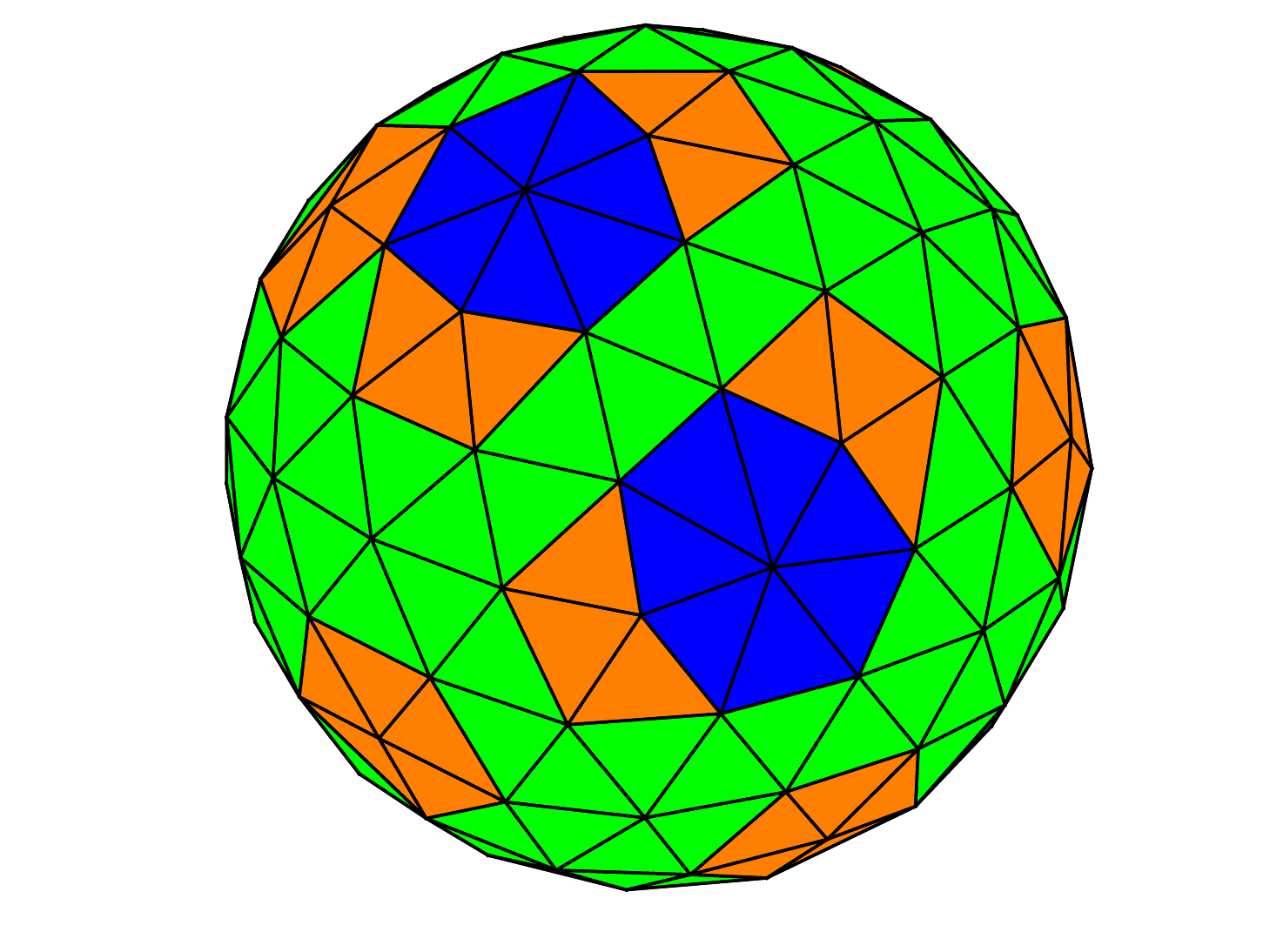}}\\
	\subcaptionbox{\label{fig:nc_N120_224}}{\includegraphics[width=0.4\textwidth]{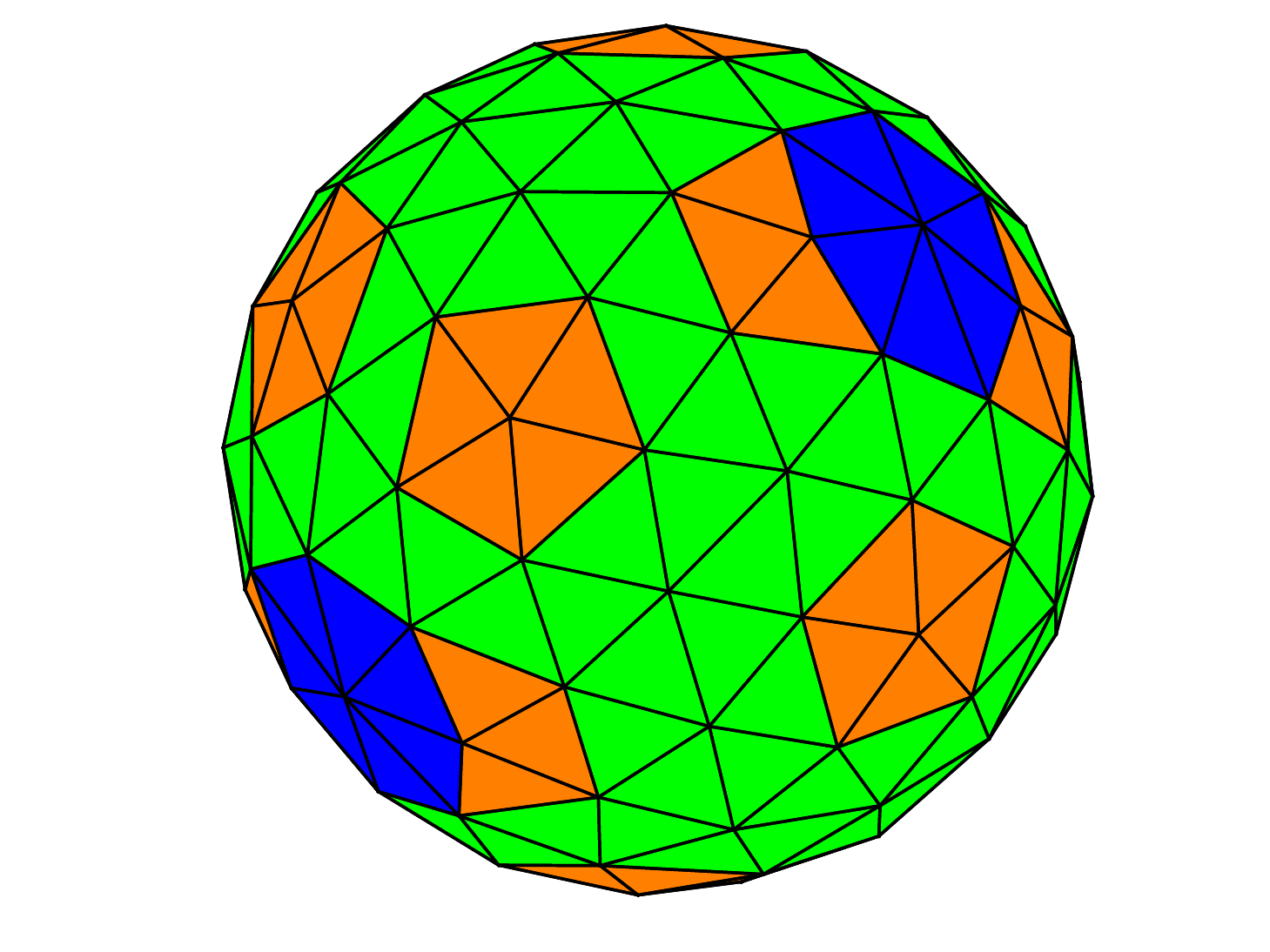}}\\
	\caption{Examples of optimal NC configurations for $N=120$ for which 238 local minima were found. \subref{fig:nc_N120_1}): The global minimum. \subref{fig:nc_N120_223}) \& \subref{fig:nc_N120_224}): Two local minima adjacent in energy.  The computed NC energies are approximately 2969.978, 2970.301, and 2970.304 respectively. (Color online).}
	\label{fig:nc_scars}
\end{figure}

\subsection{A Scaling Law for the Minimal NC Energy}

We compare the computed globally optimal energies for the NC potential with an asymptotic scaling law for equation \eqref{eqn:nc_energy} valid in the limit $N\rightarrow\infty$.  Using the same procedure as in Section \ref{sec:ne_scaling_law}, we write the NC energy as
\begin{equation}
	\mathcal{H}_\text{NC}=\mathcal{H}_\text{C}-\frac{1}{2}\mathcal{H}_\text{L}+\mathcal{H}_\text{L2},
\end{equation}
where $\mathcal{H}_\text{C}$, $\mathcal{H}_\text{L}$, and $\mathcal{H}_\text{L2}$ are defined in Eqs. \eqref{eqn:coul_energy}, \eqref{eqn:log_energy}, and \eqref{eqn:log2_energy} respectively. Substituting Eqs. \eqref{eqn:coul_scale_energy} - \eqref{eqn:log2_scale_energy} into the above yields the scaling law (also derived in \cite{lindsay2017first})
\begin{equation}
	\mathcal{H}_\text{NC}\approx \frac{N^2}{4}-\frac{1}{2}N^{3/2}+\frac{N}{8}\log N+\frac{N}{8}+\frac{1}{4}N^{1/2}-\frac{1}{24}\log N-\frac{1}{16}+\mathcal{O}(N^{-1/2}).
	\label{eqn:nc_scale_energy}
\end{equation}

\begin{figure}[h]
	\centering
	\includegraphics[width=0.7\textwidth]{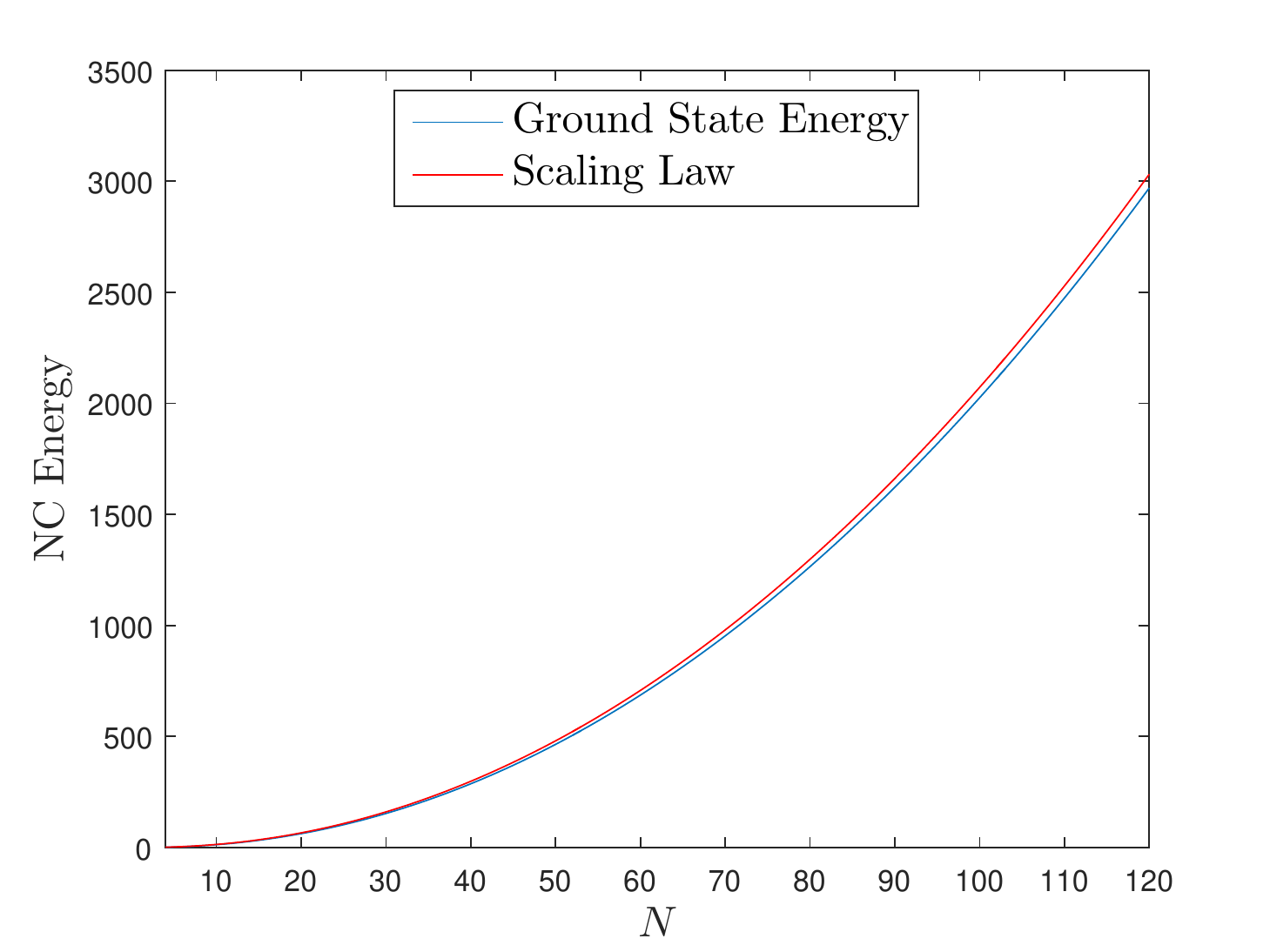}
	\caption{Comparison of globally optimal NC energies with the asymptotic scaling law \eqref{eqn:ne_scale_energy}. (Color online).   }
	\label{fig:nc_scaling}
\end{figure}

Figure \ref{fig:nc_scaling} compares the scaling law \eqref{eqn:nc_scale_energy} with the computed globally optimal energies. 

\section{Discussion}
\label{sec:discussion}

In \cite{ridgway2018iterative}, putatively optimal configurations for the Coulombic, logarithmic, and inverse-square law potentials were computed for $N\leq65$.  Non-optimal saddle points were excluded by computing the eigenvalues of the Hessian matrix.  The discovered minima were compared with previous literature \cite{erber1991equilibrium, erber1996complex, bergersen1994equilibrium} and it was found that nearly all of the optimal configurations were reproduced and that some of the previously identified minima are indeed saddle points.  Here, we use the same method to compute putatively optimal configurations of the narrow escape and narrow capture potentials for $N\leq120$.  

It is interesting that many of the local minima for all five of these potentials look qualitatively similar with respect to their scar pictures.  We compare local minima  across these potentials to determine if any are identical.  The configurations that are shared among these potentials are termed \emph{partially optimal} and these include the universal optima.  Following Section \ref{sec:numerical_computation}, we use pairwise distances to identify partially optimal configurations. In addition to the universal optima, we find that the $N=7$ and $N=32$ global minima are identical among all five potentials within a very small margin.  We note that the second of these configurations is in a special class of icosadeltahedral configurations that occur when
\begin{equation}
	N=n^2+nm+m^2 + 2, \qquad n,m\in\mathbb{Z}_+.
\end{equation}
These configurations have 12 pentagonal defects arranged at the corners of an inscribed icosahedron.  It was once believed that this class of configurations were universally optimal, however this is not the case despite the high degree of symmetry for these configurations \cite{altschuler1997possible}. The next of these arrangements occurs at $N=72$ and both the narrow escape and narrow capture potentials exhibit global minima with apparent icosadeltahedral symmetry upon visual inspection.  However, the difference in pairwise distances indicate that these configurations are in fact different. The scar structure alone is not enough to distinguish configurations since different minima for a given $N$ typically have identical coordination numbers. 

\begin{figure}[h]
	\centering
	\subcaptionbox{\label{fig:nc_N72_1}}{\includegraphics[width=0.4\textwidth]{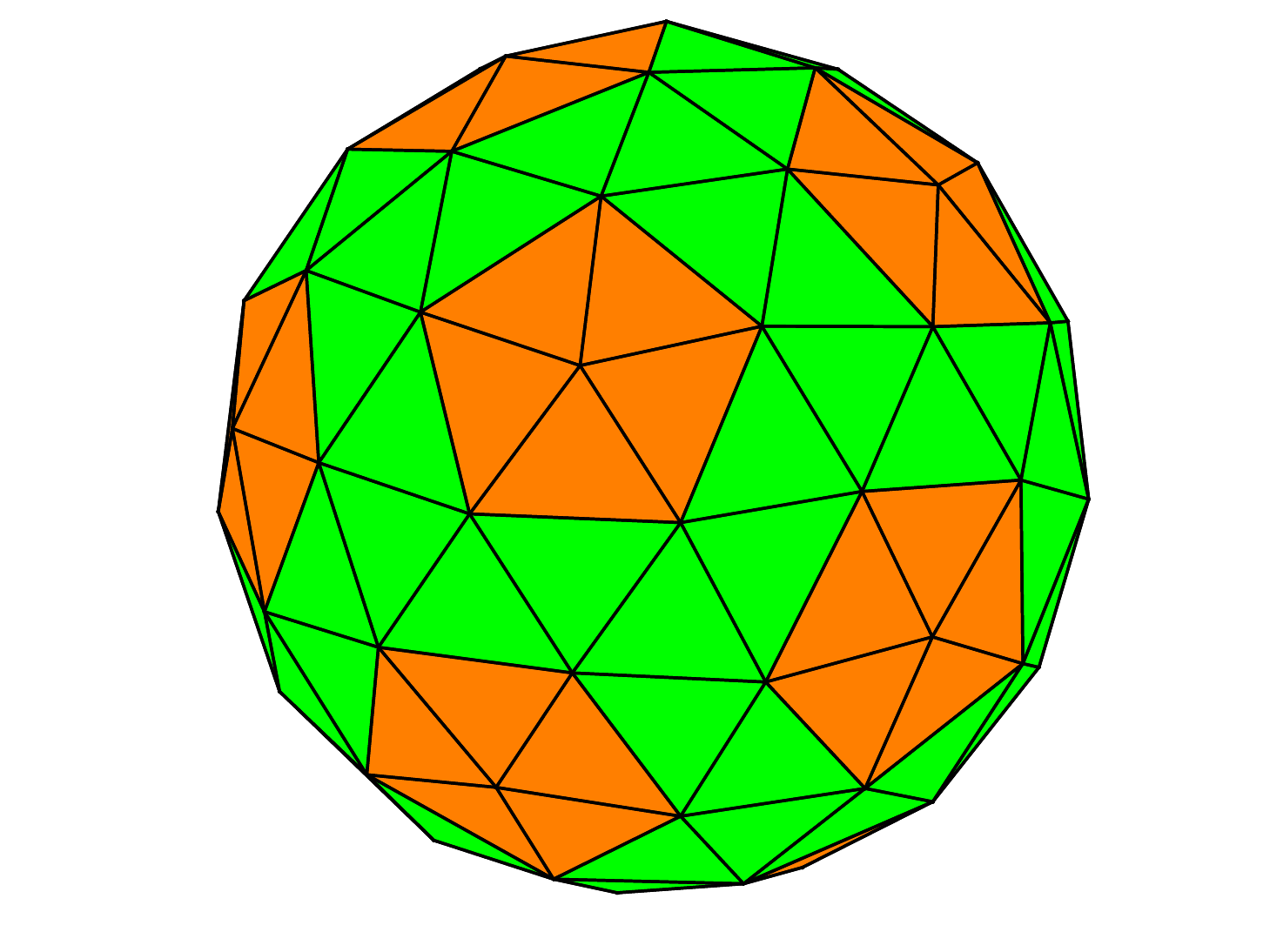}}
	~
	\subcaptionbox{\label{fig:ne_N72_1}}{\includegraphics[width=0.4\textwidth]{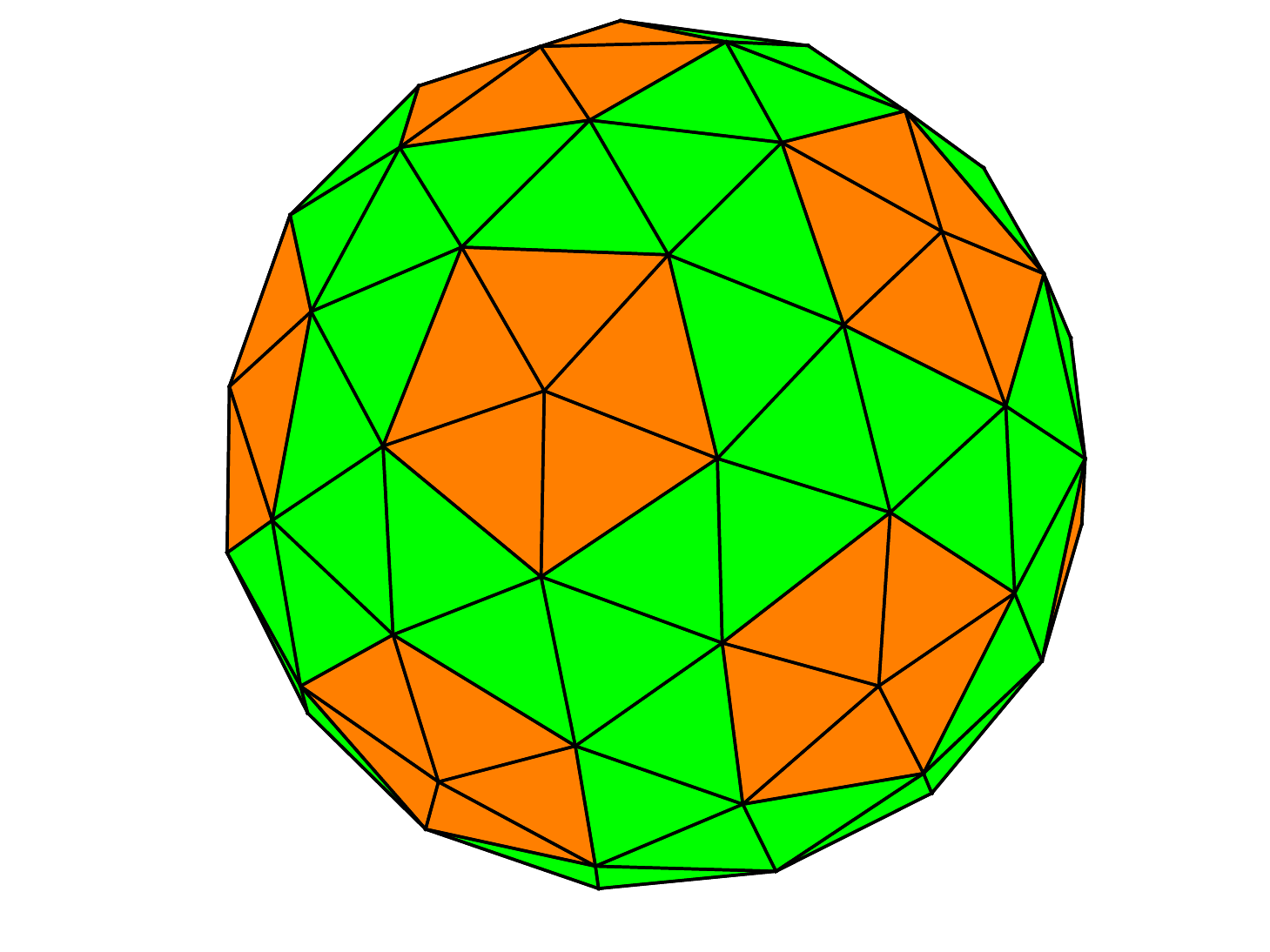}}\\
	\caption{Global minima for the NC (\subref{fig:nc_N72_1}) and NE (\subref{fig:ne_N72_1}) potentials that appear close to being partially optimal. The norm of the difference in pairwise distances is $8.67\times 10^{-3}$. (Color online).}
	\label{fig:ne_nc_comp}
\end{figure}

There are other $N$ for which the configurations are close but don't appear partially optimal.  The universal optima and the partial optima stated above agree to within a few decimal places of machine precision when compared with pairwise distances. Other configurations only agree to around $10^{-3}$ to $10^{-1}$ (e.g. when $N\lesssim20$ and $N=72$ shown in figure \ref{fig:ne_nc_comp}). 

In future work it may be of practical interest to apply results from the narrow escape problem to modelling diffusion in inverse opals. The `dwell time' (i.e. MFPT) of a diffusing particle in a single spherical cavity may be important for understanding diffusion in a connected network of such cavities. Inverse opals have been studied experimentally and numericlaly (e.g. Ref. \cite{cherdhirankorn2010tracer}) but developing a quantitative theory of diffusion is still an open problem. 


\subsubsection*{Acknowledgments}

The authors are grateful to NSERC of Canada for support through a Discovery grant and a USRA fellowship.

{

\bibliographystyle{ieeetr}
}

\begin{appendix}

\section*{Appendix: File Structure of Online Material}
\label{sec:appendix}
We provide MATLAB (.mat) files containing data on each computed local minimum and saddle point.  There are two files for each potential (NE and NC). One contains data on the minima only and the other contains data on the minima and the saddle points.  

The files with only the minima contain a $117\times4$ cell array, \verb|ALL_LEVELS|.  Each row contains data on all the local minima for a given $N$ starting at $N=4$ up to $N=120$. The data in each column contains, in order, $N$, geometric and computational data for each $N$-particle local minima, the number of $N$-particle local minima, and the number of $(N+1)$-particle starting configurations.  The data in the second column are stored in a $n$x6 cell array where $n$ is the number of local minima.  Each row in this matrix corresponds to a single local minimum and the rows are sorted in order of increasing energy.  The data in each column contains $N$, the energy of the configuration, an $N\times5$ matrix containing the particle coordinates $(\theta,\phi, x, y, z)$, the lowest energy $(N-1)$-particle local minimum from which the configuration was computed, the average number of iterations required to compute the configuration, and the number of particles in the configuration having 5 nearest neighbours (5-fold defects).   

The files with the minima and the saddle points have the same structure as described above except the second column is a $n\times8$ cell array.  The additional two columns of this array contain the smallest eigenvalue of the Hessian matrix and the $L_2$-norm of the derivative 
\begin{equation*}
	\sum_{i=1}^{N-2}\hat{\phi}_i\frac{\partial}{\partial\phi_i} +\sum_{i=1}^{N-1}\hat{\theta}_i\frac{\partial}{\partial\theta_i}
\end{equation*} 
of the pairwise energy \eqref{eqn:general_potential}. Note that two azimuthal angles and one polar angle are fixed in the above and in the Hessian computation, as explained in Section \ref{sec:numerical_computation}.

\end{appendix}

\end{document}